\documentclass[fleqn,usenatbib]{mnras}

\usepackage{newtxtext,newtxmath}
\usepackage{verbatim}

\usepackage[T1]{fontenc}
\usepackage{ae,aecompl}
\usepackage[svgnames]{xcolor}
\usepackage{hyperref}
\hypersetup{citecolor=Blue}

\usepackage{graphicx}	
\usepackage{amsmath}	
\usepackage{amssymb}	
\usepackage{amsfonts}
\usepackage{epsfig,rotating}

\def\simeq{\mathrel{\raise.3ex\hbox{$\sim$}\mkern-14mu\lower0.4ex\hbox{$-$}}}

\def\ltsima{$\; \buildrel < \over \sim \;$}
\def\simlt{\lower.5ex\hbox{\ltsima}}
\def\gtsima{$\; \buildrel > \over \sim \;$}
\def\simgt{\lower.5ex\hbox{\gtsima}}

\def\lsun{{\rm L_{\odot}}}
\def\msun{{\rm M_{\odot}}}

\def\be{\begin{equation}}
\def\ee{\end{equation}}

\def\del#1{{}}

\newcommand\mearth{\,{\rm M}_{\oplus}}
\newcommand\mj{\,{\rm M}_{\rm J}}

\title[Secondary protoplanetary discs]{The paradox of youth for ALMA planet candidates}

\author[Nayakshin]{S. Nayakshin$^{1}$\\
$^{1}$ School of Physics and Astronomy, University of Leicester, University Road, LE1 7RH Leicester, United Kingdom
}

\date{Accepted XXX. Received YYY; in original form ZZZ}

\pubyear{2019}

\begin{document}
\label{firstpage}
\pagerange{\pageref{firstpage}--\pageref{lastpage}}
\maketitle

\begin{abstract}
Recent ALMA observations indicate that the majority of bright protoplanetary discs show signatures of young moderately massive planets. I show that this result is paradoxical. The planets should evolve away from their observed states by radial migration and gas accretion in about 1\% of the system age. These systems should then hatch tens of giant planets in their lifetime, and there should exist a very large population of bright planet-less discs; none of this is observationally supported. An alternative scenario, in which the population of bright ALMA discs is dominated by secondary discs recently rejuvenated by deposition of new gas, is proposed. The data are well explained if the gaseous mass of the discs is comparable to a Jovian planet mass, and they last a small fraction of a Million years. Self-disruptions of dusty gas giant protoplanets, previously predicted in the context of the Tidal Downsizing theory of planet formation, provide a suitable mechanism for such injections of new fuel, and yield disc and planet properties commensurate with ALMA observations. If this scenario is correct, then the secondary discs have gas-to-dust ratios considerably smaller than 100, and long look ALMA and NIR/optical observations of dimmer targets should uncover dusty, not yet disrupted, gas clumps with sizes of order an AU. Alternatively, secondary discs could originate from late external deposition of gas into the system, in which case we expect widespread signatures of warped outer discs that have not yet come into alignment with the planets.

\end{abstract}

\begin{keywords}
planets and satellites: protoplanetary discs -- planets and satellites: gaseous planets -- planets and satellites: formation
\end{keywords}



\section{Introduction}\label{sec:intro}

Planet formation and protoplanetary disc evolution are deeply intertwined problems; yet we usually observe either planets around `old' stars without discs or young stars with discs but without detected planets. In 2015 ALMA revolutionised the field, showing first hints of planets immersed in discs. In this paper I argue that these observations may require a significant re-evaluation of our view of both planet formation and protoplanetary disc evolution.

A generic planet is a solid core surrounded by a volatile envelope. There exist two opposite scenarios for assembling any such planet at any separation from a host star,  and this is why planet formation remains a largely unsolved problem. In the Core Accretion (CA) scenario, planets start as tiny solid bodies and grow first by accretion of solid material and then by accretion of gas \citep{Safronov72,PollackEtal96,IdaLin04a,AlibertEtal05,MordasiniEtal09b}. Assembly of massive cores via planetesimal accretion is too slow beyond tens of AU \citep[e.g.,][]{KL99}  to result in a gas giant planet. However, \cite{OrmelKlahr10,JohansenLacerda10} pointed out that accretion of $\sim 1$~mm to $\sim 10$~cm sized pebbles could significantly speed up assembly of massive solid cores in CA. It is thus viable to hatch massive gas giants by CA even at separations of 100 AU \citep{LambrechtsJ12,LambrechtsEtal14,BitschEtal15,Mordasini18,BitschEtal19}.

In the Gravitational Instability (GI) scenario \citep{Kuiper51b},  a massive protoplanetary disc fragments onto gaseous clumps. Discs 
cannot fragment at distances closer than $\sim 50$~AU \citep{Gammie01,Rice05}, so classical GI was felt to be relevant only to the rare directly imaged planets beyond tens of AU \citep[e.g.,][]{Rafikov05,MaroisEtal10,ViganEtal17}. However, GI clumps can be delivered into the inner few AU within $\sim 10^4-10^5$ years due to gravitational interactions (migration) with a massive disc \citep{VB05,VB06,VB10,MachidaEtal10,BaruteauEtal11,MichaelEtal11,ForganRice13b,FletcherEtal19}. Sedimentation of grains within the clumps yields solid cores \citep[e.g.,][]{McCreaWilliams65,CameronEtal82,Boss98,HelledEtal08,Nayakshin10a}. Removal of some or all of the gaseous envelope then results in a very broad range of planet outcomes \citep{BoleyEtal10,NayakshinFletcher15,MullerEtal18}. This modern variant of the original \cite{Kuiper51b} scenario is now known as Tidal Downsizing \citep{Nayakshin10c,Nayakshin_Review}. As with Core Accretion, the physics of the model is diverse enough to produce planets of any mass and at any separation.

Differentiating between these diametrically opposite scenarios proved hard not only because they are complex but especially because pre-ALMA observational constraints on planet formation are indirect. Until recently we had data about (a) systems that are {\em no longer} forming planets, such as the Solar System and the exoplanetary systems \citep{FabryckyEtal14}; (b) separate to these, we have snapshots and statistical constraints on protoplanetary disc evolution and dispersion \citep{HaischEtal01,WilliamsCieza11,AlexanderREtal14a}. As rich and important as these data sets are, they fall far short of connecting planets to protoplanetary discs in a direct unambiguous way.


Recently ALMA uncovered widespread signatures of planet formation in discs orbiting $\sim 1-10$ Myr old stars on scales from 10 AU to over 100 AU \citep{BroganEtal15,AndrewsEtal16,DipierroEtal18,HendlerEtal18,LiuEtal19,Dsharp1,Dsharp2,MaciasEtal19}. These signatures are dominated by nearly circular annular rings of enhanced dust emission interspersed with gaps, some shallow, some very deep. The consensus appears to be that most of these features are carved by young planets \citep{DipierroEtal15,DipierroEtal16a,LongEtal18,DSHARP-6,Dsharp7}, although some of the features may be not due to planets \citep[e.g.,][]{ZhangEtal15-cond-front}. 

These new observations represent a quantum leap in the field and enable us to test planet formation models in new ways: this is the first time that we observe planets in the process of formation and we also have data on the properties of parent discs, such as the accretion rates onto the star, the gas molecular and the dust continuum emission, and the disc morphology. By pairing the discs with the planets explicitly rather than indirectly through statistics of disjoint data sets we can probe the {\em rates} of planet forming processes.

\begin{figure*}
\includegraphics[width=0.99\textwidth]{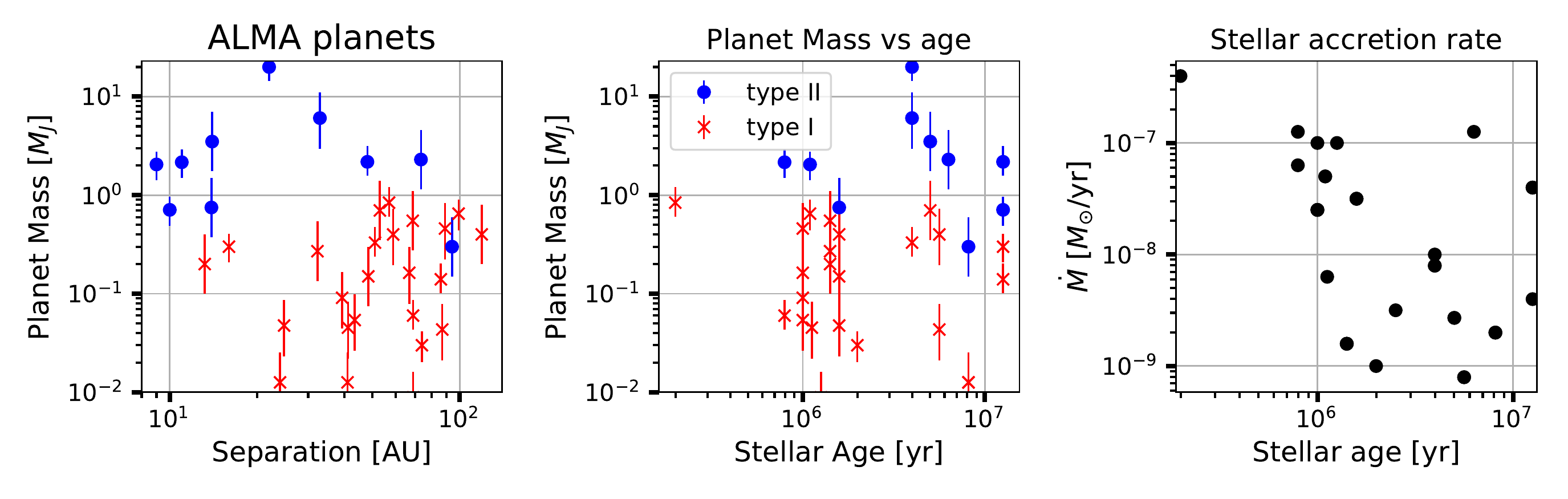}
\caption{Data used in this paper. The red crosses and blue dots show planets that we find are in type I and type II migration regime, respectively, assuming the turbulent disc viscosity parameter $\alpha_{\rm turb} = 3\times 10^{-3}$.}
\label{fig:DATA}
\end{figure*}

For example, Core Accretion models  \citep{PollackEtal96,IkomaEtal00} predict a runaway gas accretion phase during which a $\sim 1$~Neptune mass core accretes hundreds of $\mearth$ worth of gas in $\sim 10^4-10^5$~years \citep[e.g., see Figs. 1 and 2 in][]{MordasiniEtal12a}. \cite{NayakshinEtal19} showed that ALMA data require the runaway gas accretion rates to be reduced from the theoretically expected ones by about an order of magnitude since otherwise the embedded ALMA candidate planets should rapidly become gas giants with masses significantly above Jupiter for these wide separations \citep[see also][]{LodatoEtal19,NduguEtal19}. Further, \cite{Manara19-ALMA-va-PopSyn} have shown how observed gas accretion rates onto the stars can be used to check population synthesis models of planet formation.


In this paper I show that ALMA data can constrain models even further and in fact challenge most strongly not our planet formation ideas but our disc evolution and dispersion ideas. In the standard picture, protoplanetary disc removal is {\em slow} -- it takes $\sim 1 -10$~Myr, or at least thousands of orbits for the discs to go away. In \S \ref{sec:SS} we investigate this scenario and find that if it were correct for the ALMA discs then the planets must be in a very rapid state of flux, migrating into the inner disc and possibly the star within a small fraction of the system age. This paradox is present no matter how the planets formed, e.g., for both Core Accretion and Tidal Downsizing pictures. In \S \ref{sec:TD}, an alternative interpretation of the ALMA data is proposed -- that the observed discs are much lighter than expected. This is only possible if they are not in a quasi steady state of monotonic decline, but are relatively short lived ones, having been recently created by a tidal disruption of a TD pre-collapse gas giant planet. \S \ref{sec:discussion} presents the discussion of the main results of this paper. In the Appendices we  show that the other two common ways of estimating the gas mass of the disc -- by multiplying the observed dust mass by 100, or by assuming a viscous disc equilibrium with a given viscosity parameter -- lead to similar conclusions. We also show that the uncertainties in the ages of the observed systems, as large as they are, are not likely to be significant enough to invalidate our main results.

\section{Steady State Scenario: Paradox of Youth}\label{sec:SS}

Fig. \ref{fig:DATA} shows the masses and separations of the ALMA planet candidates versus the age of their star, $t_*$. The data are collected from the two major ALMA disc surveys by \cite{LongEtal18} and \cite{Dsharp1}  \citep[see][NDS19 hereafter, for detail]{NayakshinEtal19}. The right panel of fig. \ref{fig:DATA} shows  stellar accretion rates for the host stars, $\dot M_*$. The planets shown with red crosses are those that should be migrating in the type I regime (\S \ref{sec:SS}), that is, they are insufficiently massive to  perturb the gas surface density in their vicinity significantly. The planets shown with blue filled circles are those that open deep gaps in their discs (they are said migrate in the type II regime).


To predict how these candidate planets should evolve we need to know the gas disc surface density $\Sigma$ at the planet location. \cite{LodatoEtal19,NayakshinEtal19} picked `reasonable' protoplanetary disc parameters for these systems given our current understanding of protoplanetary disc evolution \citep[e.g.,][]{AlexanderREtal14a} and performed population synthesis, e.g., a statistical analysis of the problem. Here we instead use the observed stellar accretion rates to constrain $\Sigma$ more directly  for each system. There is more than one way of doing so. In \S \ref{sec:disc} we follow an observationally motivated and the least model dependent approach to link $\Sigma$ and $\dot M_*$. In the Appendix we show that two other commonly used ways of estimating $\Sigma$ lead to similar conclusions.


\subsection{Model disc setup}\label{sec:disc}

In the Steady State (SS) accretion model, we rely on the viscous disc evolution theory and calculations \citep{JPA12,RosottiEtal17} which show that the total gas disc mass at age $t_*$ is 
\begin{equation}
    M_{\rm disc} = \xi \dot M_* t_*\;,
    \label{Mdisc-ss0}
\end{equation}
where $\dot M_*$ is the accretion rate onto the host star, and $\xi \approx 2$. This equation applies after the disc reached the viscous equilibrium; at earlier times most of the mass is at radii where the viscous time is much longer than $t_*$, and calculations show that eq. \ref{Mdisc-ss0} is an under-estimate. Therefore, by using $\xi = 2$ below we follow the most conservative approach. Higher values of $\xi$ would only strengthen our conclusions. 

We assume that the disc surface density $\Sigma$ scales with $R$ as \citep[this is consistent with observations, see][]{WilliamsCieza11}
\begin{equation}
    \Sigma = \Sigma_0 \frac{R_0}{R}\;,
    \label{sigma-0}
\end{equation}{}
where $R_0 = 100$~AU is a scaling constant, out to an outer radius $R_{\rm out}$. Observations show that the molecular CO-line detected gas disc outer radii, $R_{\rm out}$, are usually a factor of $\sim 2$ larger than the mm-dust disc radii. Since ALMA candidate planets are found at separations up to 100 AU, $R_{\rm out}$ should be comparable to $R_0$ or exceed that. For simplicity we fix $R_{\rm out} = 200$~AU \citep[which is also typical for the observed CO discs; see fig. 8 in ][]{AnsdellEtal18}.

With these definitions, and the inner disc radius $R_{\rm in}  \approx R_* \sim 0$ (i.e., negligible compared to the scales of interest here), the constant $\Sigma_0$ can be found from the total disc mass $M_{\rm disc}$. For convenience we define a local disc mass at the planet orbital radius,
\begin{equation}
    M_{\rm loc} = \pi \Sigma(R) R^2 = \pi \Sigma_0 R_0 R\;.
    \label{M-loc0}
\end{equation}{}
For the disc temperature profile, we assume $T=T_0 (R_0/R)^{1/2}$,
where $T_0 = 20$~K; such temperature profiles are typical of ptotoplanetary disc models and ALMA discs \citep[e.g.,][]{LongEtal18}. The geometric aspect ratio for the discs is
\begin{equation}
    \frac{H}{R} = \left(\frac{k_b T R}{G M_* \mu} \right)^{1/2} = 0.11 \left(\frac{R}{R_0}\right)^{1/4} \left(\frac{0.7 \msun}{M_*}\right)^{1/2}\;,
    \label{HR0}
\end{equation}{}
where $H$ is the disc vertical scale height, $k_b$ is the Boltzmann's constant, $\mu = 2.45 m_p$ is the mean molecular weight, and $m_p$ is the proton mass. 


The planets are divided onto those migrating in type I (no gap in the disc) and type II (a deep gap in the disc is opened) via the \cite{CridaEtal06} criterion. The type I migration time for a planet of mass $M_{\rm p}$ is
\begin{equation}
    t_{\rm mig1} = \frac{1}{2\gamma \Omega} \frac{M_*^2}{M_{\rm p} \Sigma R^2} \left(\frac{H}{R}\right)^2 
    \label{t1-0}
\end{equation}
where $M_{\rm p}$ and $M_*$ are the planet and the star masses, $\Omega = (GM_*/R^3)^{1/2}$ is the Keplerian angular frequency. The dimensionless factor $\gamma$ depends on the disc properties \citep{PaardekooperEtal10a} and evaluates to $\gamma=2.5$ in our disc model. Numerically,
\begin{equation}
    t_{\rm mig1} = 7\times 10^{4}\; \hbox{yr}\; \left(\frac{R}{R_0} \right)^2 \left(\frac{M_*}{0.7 \msun}\right)^{1/2} \left(\frac{1 \mj}{M_{\rm p}}\right) \frac{10 \mj}{M_{\rm loc}} \;
    \label{t1-1}
\end{equation}{}

The type II migration time scale is
\begin{equation}
    t_{\rm mig2} = \frac{1}{\alpha_{\rm v} \Omega} \left(\frac{R}{H}\right)^2 \; \left(1 + \frac{M_{\rm p}}{4\pi \Sigma R^2}\right)\;,
    \label{t2-0}
\end{equation}{}
where $\alpha_{\rm v}$ is the disc viscosity parameter \citep{Shakura73}, and the last factor in the brackets corrects the migration time scale \citep{IvanovEtal99} when the planet outweighs the local disc (which is $\sim 4\pi \Sigma R^2$). This correction is usually unimportant at wide separations.

Similar to planets, dust particles drift radially inward (except possibly within the dust gap edges) due to aerodynamical friction with the gaseous disc \citep{Whipple72} at the drift velocity 
\begin{equation}
    v_{\rm dr} = \eta \left(\frac{H}{R}\right)^2 \frac{ v_{\rm K} }{St + St^{-1}}\;,
    \label{v-drift0}
\end{equation}{}
here $v_{\rm K} = (GM_*/R)^{1/2}$, and $\eta = 1.25$ given our model disc pressure profile. Eq. \ref{v-drift0} applies only far enough from the planets, e.g., at radial distance larger than a few $ R_{\rm H}$, where $R_{\rm H} = R (M_{\rm p}/3 M_*)^{1/3}$ is the Hill's radius of the planet. The dimensionless number $St$ is the particles Stokes number. Particles with $St \ll 1$ are small partilces strongly coupled to the gas, whereas those with $St \gg 1$ are larger particles weakly affected by the gas. At a disc location with gas surface density $\Sigma$, 
\begin{equation}
St = \frac{\pi}{2} \frac{a \rho_a}{ \Sigma}\;,    
\label{St-0}
\end{equation}
where we assumed the Epstein regime for a dust particle of radius $a$ and material density $\rho_a$. For definitiveness we use $\rho_a =1$~g~cm$^{-3}$ below. Particles with $St \ll 1$ are small particles strongly coupled to the gas, whereas those with $St \gg 1$ are larger particles weakly affected by the gas. 

Numerically, for $St \ll 1$, the dust drift time scale for $a=0.1$ cm particles with material density $\rho_a$ [g/cm$^3$] is 
\begin{equation}
    t_{\rm dr} = 2\times 10^{5}\; \hbox{yr}\; \left(\frac{R_0}{R} \right) \left(\frac{M_*}{0.7 \msun}\right)^{1/2} \frac{M_{\rm loc}}{10 \mj} \frac{0.1}{\rho_a a}   
    \label{t-dr0}
\end{equation}{}

\begin{figure*}
\includegraphics[width=0.99\textwidth]{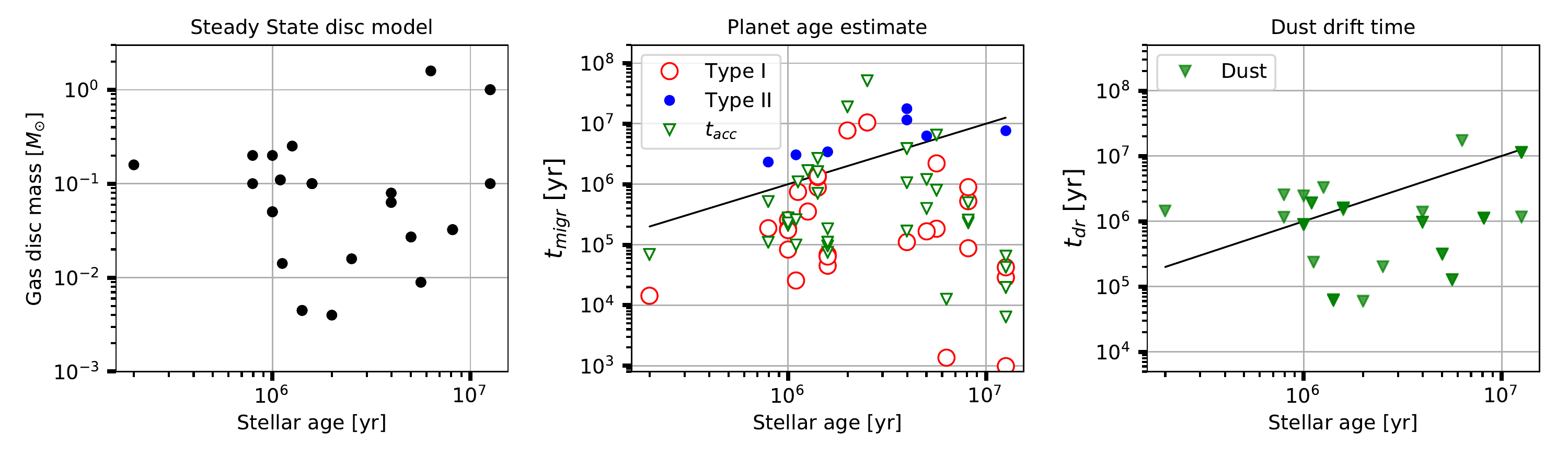}
\caption{The disc masses (left), the planet migration and accretion time scales (middle), and the dust migration time (right) for the Steady State disc and Core Accretion scenario. Many of the evolutionary time scales are much shorter than the ages of the system, constituting the paradox of youth.}
\label{fig:SS}
\end{figure*}

\subsection{Steady State disc $+$ Core Accretion}\label{sec:CA}

Now that we have our disc model specified we can start evaluating planet formation and evolution scenarios. In the Core Accretion model, planets grow in mass by accretion of solids and gas. For the high mass planets considered here, gas accretion is expected to dominate the accretion rate $\dot M_{\rm p}$ onto the planet\citep[e.g.,][]{PollackEtal96}. For definitiveness, we use the 'Bern' \citep{MordasiniEtal12a} gas accretion rate model which was found the most conservative \citep[and thus the least challenged by the ALMA data, see][]{NayakshinEtal19}. We define the planet accretion time scale $t_{\rm acc}$ as
\begin{equation}
    t_{\rm acc} = \frac{M_{\rm p}}{\dot M_{\rm p}}\;.
    \label{t-acc0}
\end{equation}{}
The planet mass should increase by the factor $e$ in time $t_{\rm acc}$.

The left panel of Fig. \ref{fig:SS} shows the total disc mass for the ALMA sample found as explained in \S \ref{sec:disc}. Before moving on to other constraints on the steady state disc scenario, it is worth pointing out that some of the disc mass estimates in Fig. \ref{fig:SS} are uncomfortably high; some of these discs should be massive enough to be gravitationally unstable. This in itself may not be a great surprise as the ALMA protoplanetary disc targets were selected to be the brightest in mm continuum sources \citep[e.g.,][]{Dsharp1}. However, two of the most massive discs are very old, $t\sim 10$~Myr. Additionally, gravitational instabilities should drive very strong spiral arm features which focus dust particles into spiral arms \citep[e.g.,][]{BoleyDurisen10}. This would contradict the dominant annular morphology of the discs in the sample. For example, the disc of the $\sim 12$~Myr old star HD169142 shows annular structures in three ALMA wavelengths, $0.89$, 1.3 and 3 mm \citep{MaciasEtal19}, yet its mass estimate in the SS model is comparable to the mass of the star. This should cause very strong spiral density features \citep{Rice05,LodatoRice05}. Therefore, already on this basis we can claim that the SS model disc masses are suspiciously high for at least some of the ALMA discs.

The middle panel of Fig. \ref{fig:SS} shows planet migration, $t_{\rm mig}$, and planet accretion, $t_{\rm acc}$, time scales. The latter are shown with green triangles. Blue filled circles and the open red circles show $t_{\rm mig}$ for planets migrating in the type II (gap opened, very massive planets) and in the type I cases (no gap, less massive planets), respectively. The results project the paradox of youth very clearly. For most of the ALMA planet candidates,  both $t_{\rm mig}$ and time $t_{\rm acc}$ are much shorter than the actual age of the system. These planets could be evolving towards very much more massive planets rapidly by runaway gas accretion, but NDS19 found that this is ruled out by the rarity of such planets at wide separations in both ALMA observations and Direct Imaging surveys \citep[e.g.][]{ViganEtal17}. The planets could then be migrating inward of 10 AU rapidly, as suggested by \cite{LodatoEtal19}. However, in that case we need a train of $\sim t_*/t_{\rm mig}$ gas giants per ALMA disc, that is, as many as $10-1000$ planets per disc. This is very unlikely because giant planets are quite rare at all separations. For example, \cite{FernandesEtal19} finds that the separation-integrated frequency of occurrence is only $\sim 6$\% for planets more massive than $1\mj$. 

The right panel of fig. \ref{fig:SS} shows that  for many of the systems the dust drift time scales are $\sim 10$ times shorter than the ages of the stars, $t_*$. This however is a very well known problem of protoplanetary discs: large dust particles drift inward much too rapidly \citep[e.g.,][]{Weiden77}. While no uniquely accepted solution exists, it is very likely that dust fragmentation is a big part of the solution. Turbulent grain fragmentation may limit dust growth so that the radial drift velocity is sufficiently small to sustain an observable dust population for Millions of years \citep{DD05,Birnstiel09,BirnstielEtal12}. A self-consistent dust growth and fragmentation modelling is beyond the scope of this paper. The point of the right panel of fig. \ref{fig:SS} is therefore to only draw attention of the modellers to the ALMA data set explored here. The previously unavailable {\em planet evolution constraints} that ALMA data provide may shed new light on the processes shaping the dust distributions in protoplanetary discs. 


 Returning to the paradox of youth of ALMA planet candidates, to illustrate it further, we propagated the planet evolution tracks forward in time for all of the planets from Fig. \ref{fig:SS}. A viscosity parameter $\alpha_{\rm v}=5\times 10^{-3}$ \citep[as in][]{LodatoEtal19,NayakshinEtal19} was assumed, and the disc surface properties frozen for this calculation. In \cite{NayakshinEtal19} it was shown that statistics of ALMA planets require slower runaway gas accretion rates, approximately by an order of magnitude, compared with the standard Core Accretion calculations. For this reason, and to also discuss other planet formation scenarios in the next section, here we compare two extreme cases: gas accretion on versus gas accretion completely off.

The results of this planet evolution calculation, carried forward for time $t=0.1$~Myr, 0.3 Myr and 0.7 Myr into the future, are shown in Fig. \ref{fig:CA-EVOL} below. The left hand panels show the calculation with the gas accretion on, whereas on the right we show the case where planetary masses kept constant. We observe that the paradox of youth is severe in both of these cases. The population of the ALMA candidate planets should evolve significantly already in $0.1$~Myr, with more than half of the planets in the $0.1 - 3\mj$ mass range disappearing from orbits wider than 10 AU. The situation becomes even worse for  the two later panels: only two candidate planets grace that parameter space. This is very uncomfortable since the ages of the stars are $\sim 1 - 10$~Myr. This calculation shows that planet migration paradox for ALMA planet candidates exists independently of whether the planet accretes gas or not. The calculation also agrees with the results of \cite{LodatoEtal19}, predicting that the planets would end up as massive planets in the future. However the challenge is that this happens all too quickly, begging the question of how can it be that we are so lucky to observe the planets in their apparently short-lived state.

\subsection{Steady State disc $+$ Tidal Downsizing}\label{sec:TD-SS}

In the TD scenario \citep{Nayakshin_Review}, gas clumps with mass of a few $\mj$ and size of an AU or more are formed very early on by GI fragmentation of very massive, $M_{\rm disc} \gtrsim (0.1-0.2) M_*$, discs. Clumps migrate inward rapidly, accreting pebbles \citep{HN18} and making massive solid cores  in their centres as they go. Many are expected to be tidally disrupted within the inner $\sim 10$~AU, perhaps leaving behind lower mass planets  \citep{Kuiper51b,BoleyEtal10,Nayakshin10c}. \cite{Nayakshin16a} showed via 1D models that lower mass dust-rich gas clumps are prone to self-disruption if they assemble massive solid cores in their centres. As the cores assemble, their accretion energy is released back into the gas envelope of the clump, inflating it. Simple energy arguments show that cores with mass $\sim 10-20\mearth$ are well capable of destabilising gas clumps with mass $M\lesssim 5 \mj$ \citep[see][]{NayakshinCha12}. Humphries \& Nayakshin (2019, submitted) have recently presented 3D simulations of dust and core dynamics within such self-inflated gas clumps, confirming results of \cite{Nayakshin16a}. Thus, TD scenario produces planets with masses and separations consistent with the ALMA planet candidates naturally and potentially very rapidly.

\begin{figure*}
\includegraphics[width=0.45\textwidth]{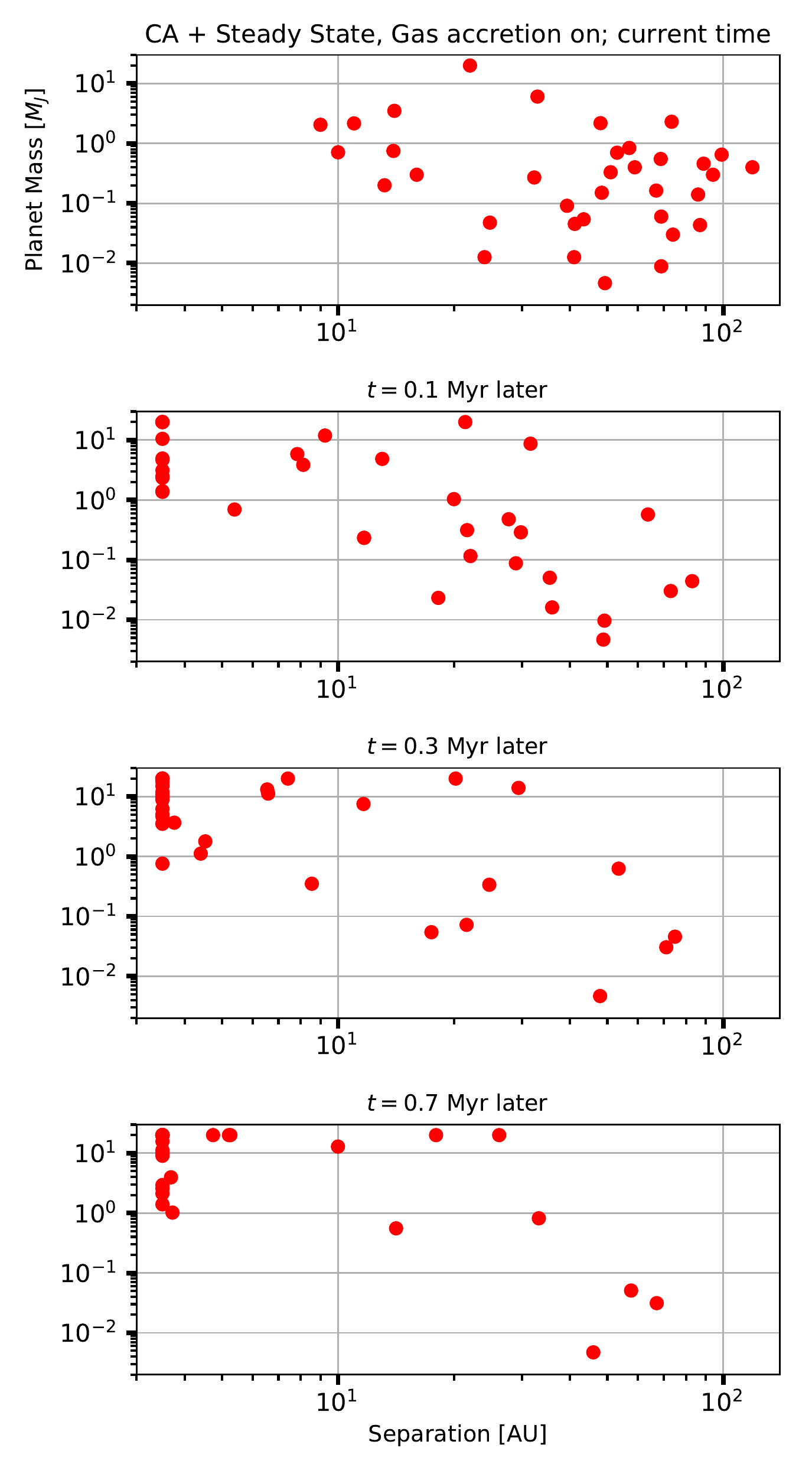}
\includegraphics[width=0.45\textwidth]{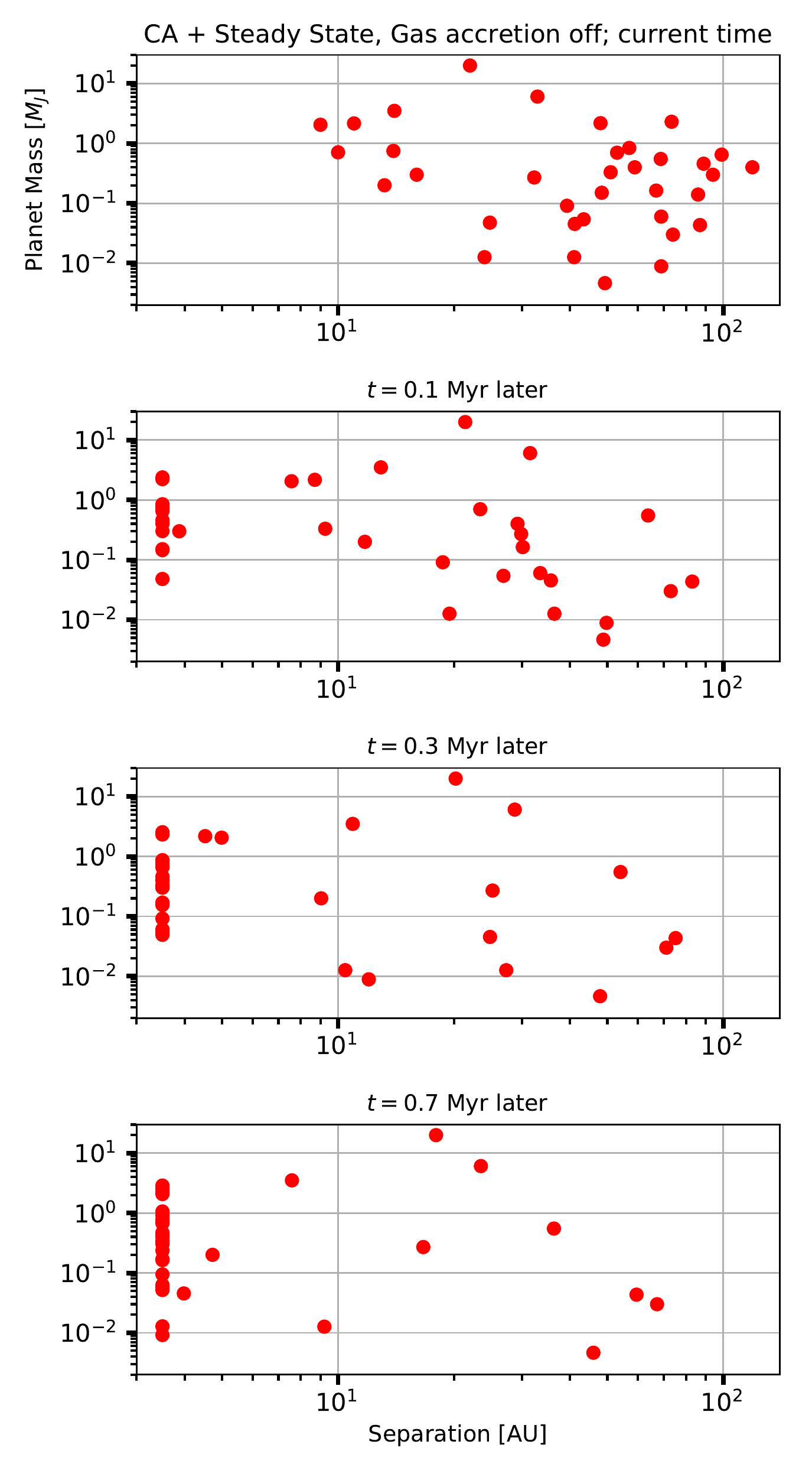}
\caption{Evolution of the ALMA candidate planets in the Steady State Core Accretion scenario with gas accretion on (left panels) or off (right panels). In both cases the planets evolve very rapidly, confirming the paradox of youth. Injection of very massive cores every $\sim 0.1-0.2$~Myr is needed to account for the ALMA observations.}
\label{fig:CA-EVOL}
\end{figure*}

However,  in conjunction with the standard assumptions about disc evolution that we explore in this section -- quasi-steady state, monotonically evolving discs -- the TD scenario is even less promising that CA as an explanation for the origin of the ALMA planet candidates. This is because the clumps are born very early (at $t\sim 0.1$~Myr) in discs that were presumably even more massive than they are now. Further, the planets are also more massive at that time in this scenario. In the Type I migration these planets must migrate inward at the rate proportional to the product $M_{\rm p} M_{\rm loc}$. Re-scaling eq. \ref{t1-1} for $M_{\rm loc}\sim 0.1 \msun$ we get
\begin{equation}
    t_{\rm mig1} \sim 10^4\hbox{ yr}\;.
    \label{t1-TD0}
\end{equation}{}
Such short time scales are indeed found in numerical simulations \citep[e.g.,][]{BoleyEtal10,BaruteauEtal11,HallCEtal17,FletcherEtal19}. Therefore, TD-made planets should migrate inward even faster than the planets in the 'migration off' Core Accretion model presented in the right panels of Fig. \ref{fig:CA-EVOL}. We should discard the `Steady State disc $+$ TD' scenario without spending much more time on it.

\subsection{Discussion of the Paradox of Youth}\label{sec:discussion-of-youth}

We found that in the Steady State disc evolution scenario, the ALMA planets must have just been made (e.g., the minimum of $t_{\rm mig}$ and $t_{\rm acc}$ ago) for us to stand a decent chance of observing them. This is true for both Core Accretion and Tidal Downsizing theories of planet formation but is highly unlikely for statistical reasons, as gas giant planets are quite rare at all separations. Furhtermore, some of the systems are inferred to host multiple planets, usually with a wide range of planet masses and separations. This would be a tremendous coincidence if these planets grew/evolved at just the right rates at very disparate conditions for us  to observe them just before they plunge into the inner disc. A case in point is CI Tau \citep{ClarkeEtal18}. In addition to the three planets at tens of AU separations analysed here, it also hosts a very massive $M_{\rm p}\approx 12\mj$ hot Jupiter at separation of only $\sim 0.1$~AU.  The planet luminosity is consistent with the hot start models \citep{FlaggEtal19}, potentially implying the planet was made by GI and migrated to its present location. Being as massive as it is, the planet should be migrating in the type II regime, with $t_{\rm mig2} \approx M_{\rm p}/\dot M_* \sim 3\times 10^{5}$~yr
\begin{equation}
    t_{\rm mig2} \approx \frac{M_{\rm p}}{\dot M_*} \sim 3\times 10^{5}\hbox{ yr}\;   \label{tmig2-CItau}
\end{equation}{}
for the accretion rate of $3\times 10^{-8}\msun$~yr$^{-1}$. The migration times of the outer three planets in this system are evaluated in the SS scenario as $\sim 7\times 10^4$ yrs for the inner two planets, and $\sim 3$~Myr for the outermost planet. For a system that is $\sim 1-2$~Myr old, $t_{\rm mig}$ is uncomfortably short for the inner 3 planets.

In Appendix \ref{sec:age-un} we consider how the very large age uncertainties of the stars younger than $\sim 20$~Myr \citep{SoderblomEtal14} could affect our conclusions.


\section{Tidal Downsizing: Secondary Discs}\label{sec:TD}


\subsection{Motivation}\label{sec:motiv}

We have argued that in the Steady State scenario for disc evolution, the ALMA planet candidates must be much younger than their stars for us to observe them, and that this scenario is therefore unlikely to be correct. We now consider an alternative explanation for what these observations imply. 

The most difficult challenge we face is the rapid migration of moderately massive planets, $M_{\rm p}\lesssim (0.1-1)\mj$, that are in the Type I migration regime. The migration time of such planets is proportional to the local mass of the disc, $M_{\rm loc}$ (see eq. \ref{t1-1}). Setting this mass to $M_{\rm loc} = 1 \mj$ for the moment, we get for $M_{\rm p} = 1\mj$
\begin{equation}
    t_{\rm mig1} \sim 10^6 \hbox{ yrs}\;.
    \label{t1-2}
\end{equation}{}
This is much more palatable, however this would imply that the disc cannot be a very long lived one. Taking accretion rate onto the star to be $\dot M_* =10^{-8} \msun$~yr$^{-1}$ we see that the disc would be completely consumed by the star within the depletion time defined as
\begin{equation}
    t_{\rm dep} = \frac{M_{\rm disc}}{\dot M_*} \sim 10^{5} \hbox{ yr }\;
    \frac{M_{\rm disc}}{1\mj} \; \frac{10^{-8}\msun \hbox{ yr}^{-1}}{\dot M_*}\;.
    \label{tdep0}
\end{equation}{}

This time is quite short. In the Steady State scenario, such a situation could arise if the discs used to be much more massive and we just happen to observe them just before they are completely dispersed. This solution would imply that there should be many other discs that are much more massive than those in the ALMA sample since they are not yet close to dispersion. This is not the case; the discs in our sample are some of the most massive, and have on average an order of magnitude more dust than a typical protoplanetary disc of the same age \citep[e.g., see Fig. 4 in][]{LongEtal18}. This is not a coincidence as so far ALMA focused on targets that are the brightest in the mm-wavelength emission \citep[e.g.,][]{Dsharp1}. 

Therefore, if we are prepared to contemplate that the discs in this sample are surprisingly low mass, e.g., $\sim 1\mj$ in gas, then we also need to accept that they are transient, that is much shorter lived than the age of the systems. If this is the case then we do not have to require planet migration time scales to be as long as the age of the star -- they only have to be longer than the disc depletion time scale. 

\subsection{Secondary discs in Tidal Downsizing}\label{sec:TD-connection}

In the Core Accretion scenarios, planets accrete planetesimals, pebbles, and gas from the parent disc \citep[e.g.,][]{Mordasini18} while the disc is evolving monotonically towards its dissipation via accretion onto the star and also through outflows \citep{AlexanderREtal14a}. Hence the flow of mass is always from the disc to the planet.

In Tidal Downsizing, mass flows between planets and the disc in both directions. Planets gain mass from the disc at their formation via gravitational instability, of course. They may continue to accrete gas \citep{KratterEtal10,ZhuEtal12a,Stamatellos15} but only if gas cooling in the Hill sphere of the planet is rapid enough \citep{Nayakshin17a}. They also accrete both planetesimals \citep{HelledEtal06} and pebbles \citep{BoleyDurisen10,BoleyEtal11a,Nayakshin15a,HN18,Forgan19}. 

TD planets return some or all of their gaseous mass to the disc when they are tidally disrupted \citep{VB05,VB06,BoleyEtal10,MachidaEtal10}. This disruption process is a promising candidate to explain the FU Ori outbursts of very young ($t \lesssim 0.5$~Myr) protostars \citep{VB10} and associated puzzles, such as the `lumonisity problem' of young protostars \citep{DunhamVorobyov12}, and the age spread of accreting stars in clusters \citep{BaraffeEtal12}. TD protoplanets are disrupted in this case in the inner $\sim 0.1$ AU to a few AU \citep{NayakshinLodato12}. Right after their disruption, the local disc is awash with material formerly belonging to the planet, which fuels a powerful gas accretion outburst onto the star. The outburst lasts on the order of the local disc viscous time, 
\begin{equation}
    t_{\rm visc} = \frac{1}{\alpha_{\rm v} \Omega} \frac{R^2}{H^2} \sim 60\; \hbox{yr } \alpha_{-1}^{-1} R_{-1}^{3/2}\;,
    \label{tvisc0}
\end{equation}{}
where $\alpha_{-1} =\alpha_{\rm v}/0.1$ and $R_{-1} = R/(0.1$~AU), and we assumed $H/R = 0.03$, as appropriate for the inner disc region. The duration of these outburst can hence range from $\sim 10$~yrs to thousands of years, and the corresponding accretion rate onto the star, assuming that the planet looses $3\mj$ of gas, is
\begin{equation}
    \dot M_* \sim \frac{M_{\rm p}}{t_{\rm visc}} \sim 3\times 10^{-5} \frac{\msun}{\hbox{yr}}\frac{100 \hbox{ yr}}{t_{\rm visc}}\;.
    \label{dotM0}
\end{equation}{}
This is comparable to the stellar accretion rates during FU Ori outbursts \citep{HK96} but is multiple orders of magnitude larger than the gas accretion rates we observe for the ALMA sample (see the right panel in Fig. \ref{fig:DATA}).

However, as stressed previously, TD planets can be disrupted also at tens of AU if their envelopes get over-inflated due to the energy release by the massive solid cores growing in the planet centres \citep{Nayakshin16a}, Humphries \& Nayakshin (2019, submitted). Such disruptions may release not only gas but also small dust and planetesimal like materials \citep{ChaNayakshin11a,NayakshinCha12}. Since the protoplanets can be much hotter than the surrounding gaseous discs (from hundreds up to $\sim 2000$ Kelvin), the solids released back into the disc may carry crystalline and other thermally-processed materials \citep{Vorobyov11,IleeEtal17}, which may help to account for the unexpectedly complex composition of comets in the Solar System \citep{BridgesEtal12a}. The expected duration of the outburst in this case is obtained by rescaling equation \ref{tvisc0}, assuming $H/R \sim 0.1$ and $R\sim 100$~AU:
\begin{equation}
    t_{\rm visc} \sim 1.6 \times 10^5 \; \hbox{yr } \alpha_{-1}^{-1} R_{100}^{3/2}\;,
    \label{tvisc100}
\end{equation}{}
where $R_{100} = R/(100$~AU). The corresponding accretion rate, again using $M_{\rm p} = 3\mj$ is
\begin{equation}
    \dot M_*  \sim 2\times 10^{-8} \frac{\msun}{\hbox{yr}}\frac{1.6\times 10^5 \hbox{ yr}}{t_{\rm visc}}\;.
    \label{dotM100}
\end{equation}{}
Such accretion rates are reasonably large for the ALMA data set. Furthermore, if ALMA planets are to be interpreted in the context of Tidal Downsizing, then disruption of the primordial gas clumps is a necessity. The  clumps formed by disc fragmentation due to gravitational instability are expected to be on the order of $1\mj$ to as much as $10 \mj$ given the uncertainties in the disc parameters at fragmentation \citep{KratterEtal10,BoleyEtal10,ForganRice11,KratterL16}. It does not appear likely that disc fragmentation could hatch gas clumps as low mass as $0.1\mj$. Therefore, by invoking tidal disruption of gas protoplanets with masses of a few $\mj$ via energy released by massive cores inside these protoplanets we may achieve several desirable consequences at once.

\subsection{ALMA data in the TD interpretation}\label{sec:TD-vs-ALMA}

As explained earlier, we now assume that the ALMA planet candidates are the remnants left over from the disruption of an earlier more massive progenitors\footnote{Since most of the planetary candidates are less massive than a Jovian mass \citep{LodatoEtal19,NayakshinEtal19}, this is a reasonable assumption. The few really massive planets like PDS 70 ($M_{\rm p} \sim 10 \mj$), for which this assumption is unlikely to be correct, do not actually matter much in our analysis of the paradox of youth. These massive planets easily open a deep gap in the disc and migrate inward quite slowly in the Type II regime. None of them are problematic (see the blue circles in Fig. \ref{fig:SS}).}. In this scenario most of ALMA discs are the secondary discs formed by the same disruption. This is in difference to the `primary' protoplanetary discs which we speculate have been dispersed very early on (see \S \ref{sec:example} for a more detailed calculation). 

For this section, the orbital location of the planet then sets the disc `age' via the viscous time (eq. \ref{tvisc100}). As we do not know $\alpha_{\rm v}$ a priory, we consider three representative values for it, $\alpha_{\rm v} = 0.001$, $0.01$ and $0.1$. Our calculations of planet migration and dust drift time scales in this section are identical to that in \S \ref{sec:CA} in every aspect except for how the gas surface density at the location of the planet found. 

To that end we assume that the disc reached a local viscous equilibrium everywhere inward from the location of the planet \citep{Shakura73}, which gives 
\begin{equation}
    \dot M_* = \dot M(R) = 3\pi \alpha_{\rm v} c_s H \Sigma \quad \hbox{ for any } R \le R_{\rm pl}\;,
    \label{dotM-TD}
\end{equation}{}
where $\dot M(R)$ is the accretion rate through the disc at radius $R$, and $R_{\rm pl}$ is the orbital separation of the planet from the star. Applying this equation at the location of the planet we get $\Sigma$ at that location. Further, with the chosen disc temperature profile (\S \ref{sec:SS}), the resultant disc surface density has the same power-law radius dependence as in eq. \ref{sigma-0}, that is, $\Sigma(R) \propto R^{-1}$. The mass of the disc from $R=R_* \approx 0$ to $R_{\rm pl}$ integrates to
\begin{equation}
    M_{\rm disc} = \pi \Sigma(R_{\rm pl}) R_{\rm pl}^2 = 
    \frac{1}{3} \dot M_* t_{\rm visc}\;,
    \label{Mdisc-TD0}
\end{equation}{}

\begin{figure*}
\includegraphics[width=0.8\textwidth]{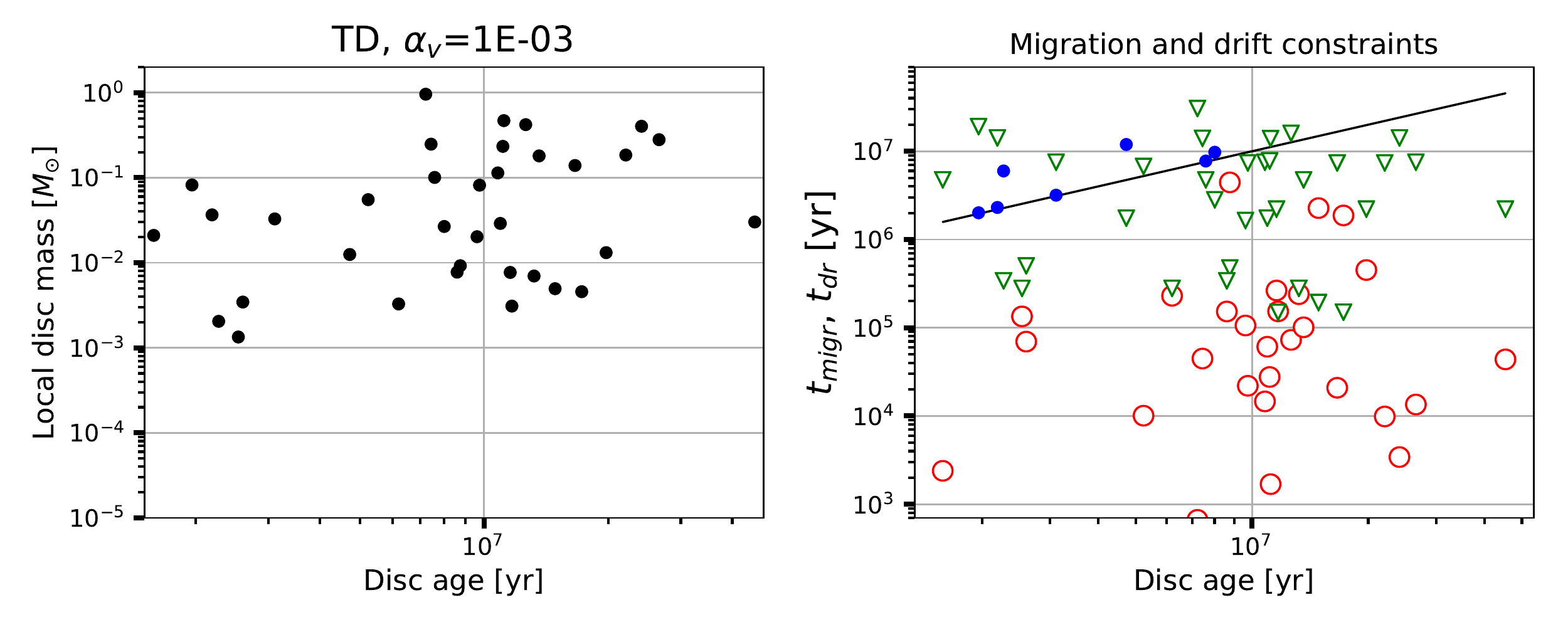}
\includegraphics[width=0.8\textwidth]{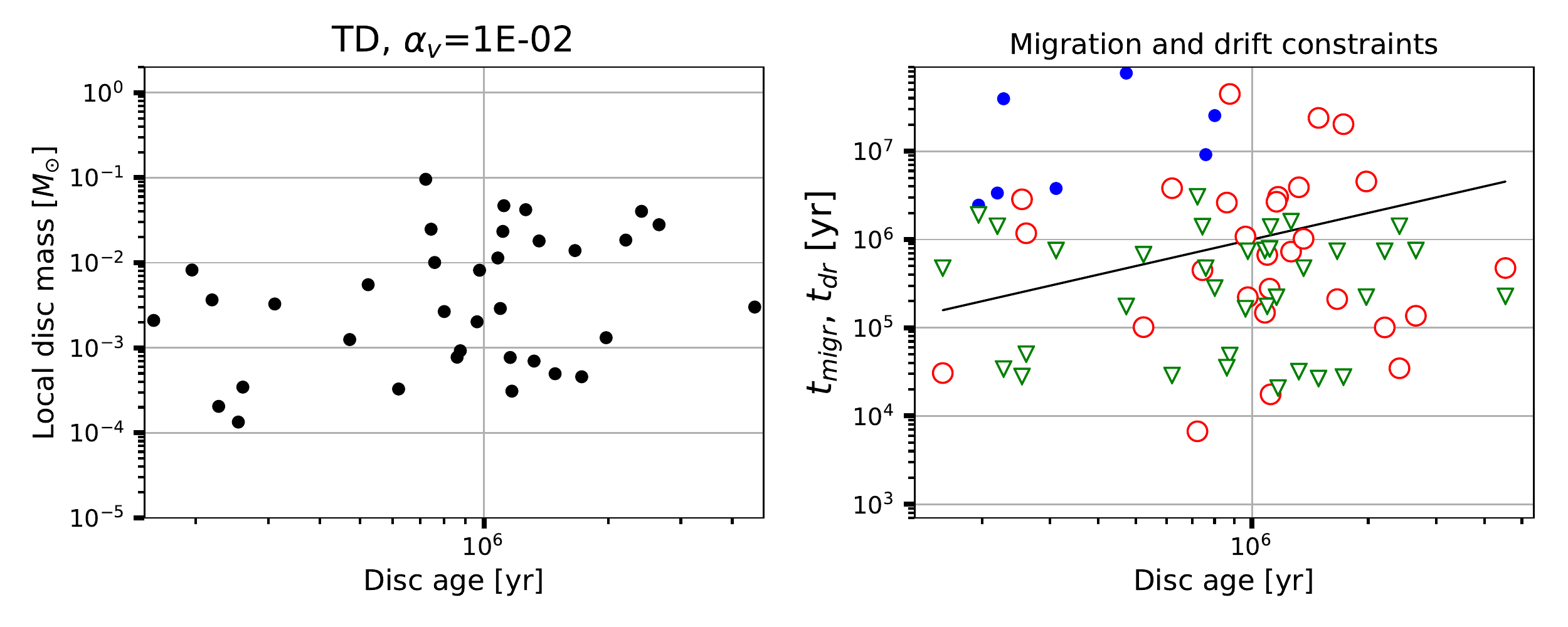}
\includegraphics[width=0.8\textwidth]{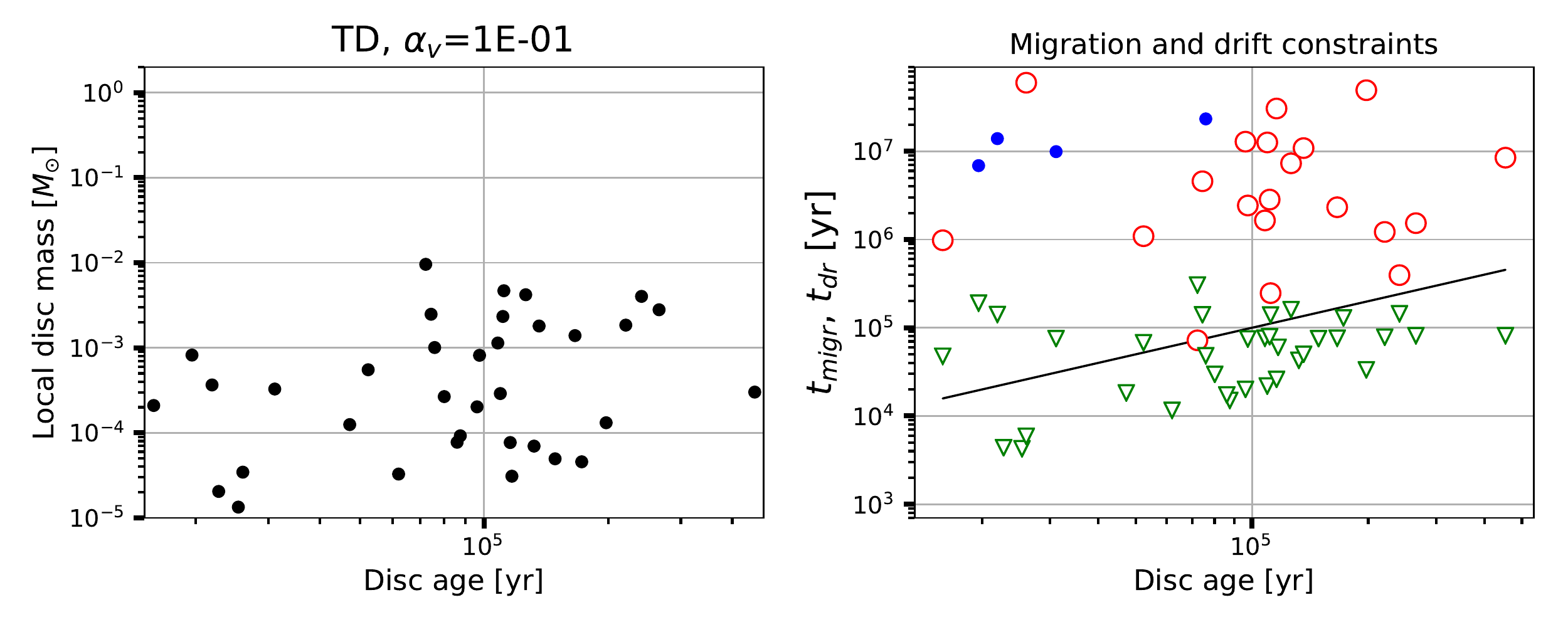}
\caption{Disc mass (left), planet migration and dust drift time scales (right) for the Tidal Disruption scenario. For $\alpha \lesssim 10^{-3}$ the disc age would be unreasonably long; for $\alpha = 10^{-2}$ (middle row panels), migration and dust time scales are too short, and the disc mass exceeds what a gas giant planet can deliver; the model with $\alpha = 0.1$ contradicts none of the three constraints.}
\label{fig:TD-migr}
\end{figure*}

Fig. \ref{fig:TD-migr} shows application of this `rejuvenated disc' scenario to the ALMA data. The left panels show the estimated gas disc masses (eq. \ref{Mdisc-TD0}), while the planet migration and the dust drift times are shown in the right panels. These are shown against the viscous time $t_{\rm visc}$ (eq. \ref{tvisc100}) rather than the system age. This is because we assume that we observe the system in a special time -- after the planet disruption -- and the state of the system depends on the age of the star very weakly. The dust radial migration time scale is therefore a very significant constraint on this scenario -- if the dust migrates inward too rapidly then these discs would not be observable via dust continuum emission. There are also no planet accretion time scales to report in this scenario. We assume that these planets do not accrete gas from the surrounding disc since their cores are very luminous; if anything these remnant planets may be loosing more mass as time goes on.

The top row of fig. \ref{fig:TD-migr} shows that for $\alpha_{\rm v} = 10^{-3}$ the discs would actually be very long-lived, requiring disc masses orders of magnitude higher than a gas giant disruption can deliver. The planets would be drained onto the star too rapidly. For many of the ALMA systems the estimated viscous times are actually longer than the stellar age, $t_*$, indicating that our assumption of viscous quasi-equilibrium (eq. \ref{dotM-TD}) is violated. This contradicts observations directly because the gas liberated in a planet disruption would not yet have reached the star since $t_{\rm visc} \ll t_*$, so this model could not explain gas accretion onto the star. Therefore, the scenario with $\alpha_{\rm v} = 10^{-3}$, and any lower values of the viscosity parameter, fail on many accounts.

The middle row in fig. \ref{fig:TD-migr} indicates that the model with the higher value of $\alpha_{\rm v} = 0.01$ does better but is still ruled out for similar reasons. 

The model with the highest value $\alpha_{\rm v} = 0.1$ (the bottom row) appears to be reasonable: the disc masses are all in the planetary regime, the planets are safe from falling into the star during the short secondary disc existence, and the dust drift time scales are somewhat shorter than $t_{\rm v}$, allowing the dust to accumulate into the planetary traps, but not so short than we expect only very narrow dust rings.

\subsection{A disc-planet co-evolution example}\label{sec:example}

Having confirmed in the previous section that the secondary disc model does much better in addressing the ALMA data than the Steady State scenario, we now present an example detailed calculation of the planet-disc co-evolution in the TD framework. For simplicity we assume just one progenitor planet in the disc.

Our calculation is based on the 1D viscous time-dependent disc evolution model with a planet embedded in it \citep{Nayakshin15c}. The disc evolves under the action of viscous and planet migration torques, and photo-evaporation. Additionally, if/when the planet is tidally disrupted  its gaseous envelope is deposited back into the disc at the planet location. The calculation includes a module for the internal evolution of the planet, which treats radiative cooling of the planet; growth, settling and vaporisation for four different species of dust (from water ice to Fe), and core formation. The accretion energy of the core is released back into the gaseous envelope. The most recent summary of the code is provided in Appendices A2-A4 of \cite{Nayakshin16a} (N16 hereafter).

We first of all note that planetary disruptions with properties similar to those that we require for the ALMA data were already seen in N16. For example, Fig. 2 in N16 shows a gas clump injected into the disc at separation of 80 AU. The clump migrates in, and gets disrupted (thick curves in that figure) at separation of about 50 AU at time $0.2$~Myr after its injection into the disc. The disruption occurs only because of the energy released by the growing core, which reaches the mass of about $12 \mearth$. The calculation with the core feedback turned off (thin curves in fig. 2 in N16) does not disrupt the clump: in that case it continues to migrate and becomes a gas giant planet that stops migrating at about 10 AU when the disc is finally dispersed. In contrast, the calculation with the core feedback produces a massive core with a small Hydrogen envelope that gets stranded at separation of 44 AU.

Therefore, we use the code of N16 with very small modifications. First of all, we modify the initial conditions. For this paper we are interested in a case when a gas clump remains at large separations at the end of the primary protoplanetary disc dissipation. Physically, clumps can be left stranded at wide separations if the disc over-extends itself during GI fragmentation, making too many clumps; such a gravitational fragmentation catastrophe is possible behind $R \gtrsim 100$~AU where disc cooling is sufficiently rapid \citep{Gammie01}. The migration of the clumps could then be much slower than that obtained when considering just one clump because the combined mass of the clumps could exceed that of the disc at the region that underwent fragmentation. Further, some clumps can be scattered outward and then have unusually slow migration histories.  

Bearing that in mind, we start our calculation close to the time of primary disc dissipation, assuming the remaining disc mass is $M_{\rm disc} = 10 \mj$. The mass of the star is set to $1\msun$. We use the $\alpha_{\rm v} = 0.1$ viscosity parameter that was seen best to explain the data in \S \ref{sec:TD-vs-ALMA}. Due to this, the primary disc is drained very rapidly by accretion onto the star. We are thus justified in setting $t\approx 0$ as the initial time in our calculation. The disc is endowed with twice the Solar metallicity pebble complement ($Z = 2 Z_\odot$), and the pebble size is set to $a=1$~mm. The protoplanetary disc surface density is set initially to
\begin{equation}
    \Sigma(R) \propto \;\frac{1}{R} \;\exp\left[ -\frac{R}{R_{\rm exp}}\right]\;,
\end{equation}{}
where $R_{\rm exp} = 150$~AU.
An $M_{\rm p} = 2\mj$ gas clump is embedded in the disc at $a=120$~AU at $t=0$. The planet is assumed to have accreted pebbles previously and has metallicity twice that of the parent disc, so $4 Z_\odot$, at the start of the calculation. This is physically feasible: \cite{HN18} found that pebbles can be accreted by a few Jupiter mass gas clump very rapidly; unlike the case of much less massive planets, no 'pebble isolation gap' \citep{LambrechtsEtal14} is opened in the initial massive disc while the clump migrates rapidly.

The only other change to the code of N16 is the addition of the stellar irradiation heating of the clump. As described in \cite{Nayakshin15c}, external clump irradiation retards its radiative cooling. While the clump is embedded in the disc, the irradiation temperature is given by the disc midplane temperature, with the same profile as used in \S \ref{sec:disc}. Once the primary disc dissipates, however, stellar radiation reaches the planet directly. The equilibrium temperature, that we call the `irradiation temperature', is the temperature reached by the surface of a passively irradiated planet, and it is given by
\begin{equation}
    T_{\rm irr} = \left[\frac{L_*}{16 \pi \sigma_B R^2}\right]^{1/4} 
\end{equation}
where the protostellar luminosity is set to $L_* = 2\lsun$. 

\begin{figure*}
\includegraphics[width=0.99\textwidth]{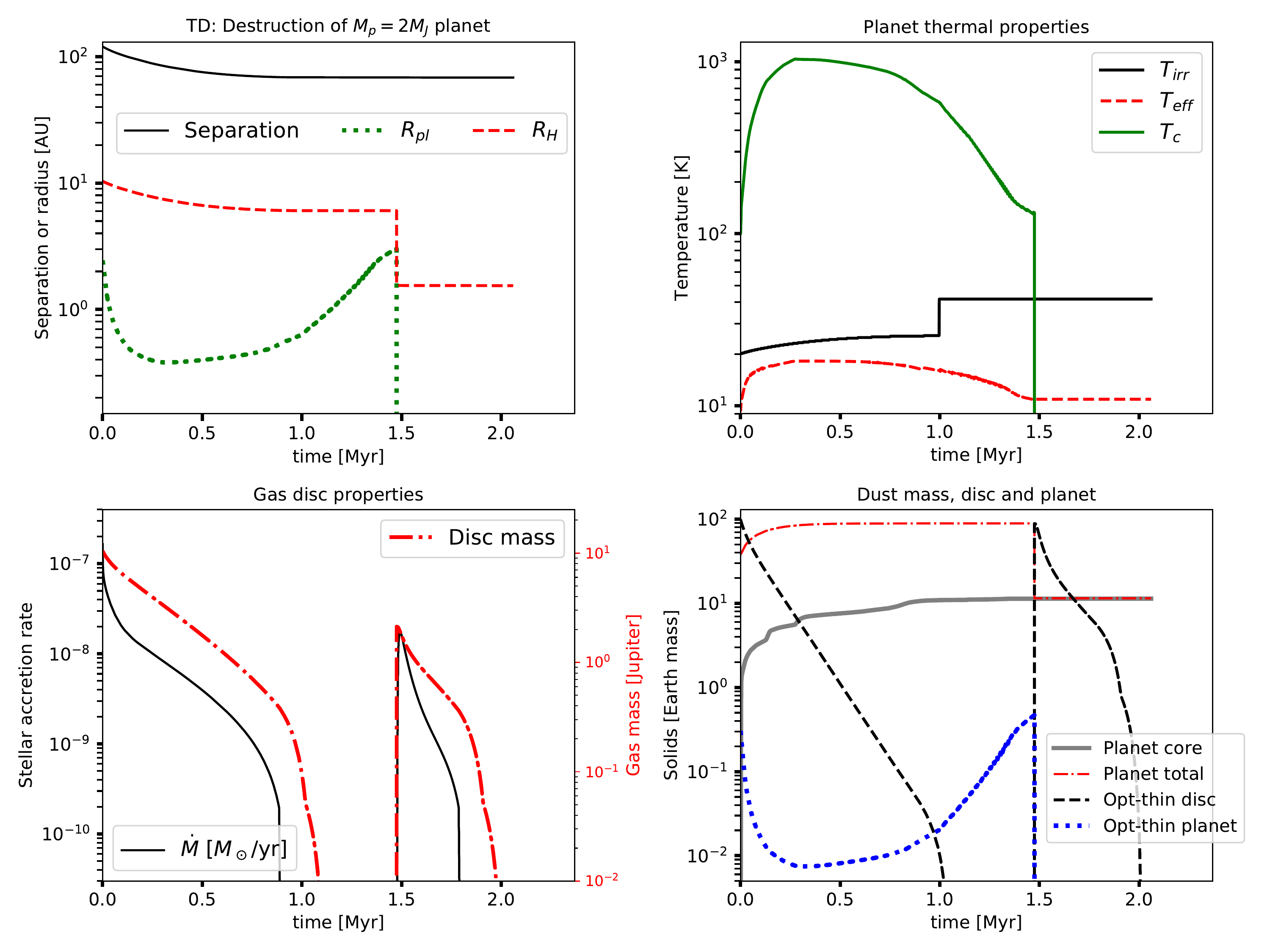}
\caption{An example disc and planet co-evolution in the Tidal Downsizing scenario. The primary disc is dissipated at around $t=1$~Myr, and the short lived secondary disc is born when the planet is disrupted at around 1.5 Myr. See \S \ref{sec:example} for more detail.}
\label{fig:TD-Secondary-Disc}
\end{figure*}

The results of this disc-planet co-evolution calculation are presented in fig. \ref{fig:TD-Secondary-Disc}. The upper left panel shows the separation $R$, planet radius $R_{\rm pl}$ and its Hill radius $R_{\rm H}$. The planet gaseous envelope is assumed to be entirely removed when $R_{\rm pl}$ exceeds\footnote{In reality the envelope may start loosing mass earlier due to thermal escape via the L1 point primarily \citep[see][]{NayakshinLodato12} before it fulfils the complete disruption criterion. The mass loss rate from the planet also depends on its spin and 3D morphology \citep[e.g.,][]{BoleyEtal10,GalvagniMayer14}. Finally, some or even much of the envelope may remain bound. We do not model these intricate effects here.} $R_{\rm H}/2$. The upper right panel shows the planet central temperature $T_{\rm c}$, the irradiation temperature $T_{\rm irr}$ and the effective temperature for the planet outgoing radiation. The bottom left panel shows the disc gas accretion rate onto the star (black solid) and the total disc mass (red dotted). The bottom right panel shows the disc and the planet dust properties, which we explain in greater detail below.

The disc is drained onto the star and is also photo-evaporated within the first 1 Myr. As the disc photo-evaporation is most efficient at relatively small radii \citep{AA09}, $R\sim 1$~AU, accretion onto the star terminates before the outer disc regions are removed. The planet manages to migrate to separation $R\approx 68$~AU before the disc is removed. While the disc is present, the planet accretes more pebbles and its total dust mass increases from $\sim 40\mearth$ to $\sim 80\mearth$ (cf. the red dot-dash curve in the bottom right panel). The core in the planet grows initially very rapidly from a negligible value to a few Earth masses at which point its growth slows down because the core luminosity starts to affect the contraction of the planet and the regions nearest to the core. Due to this the rate of the planet central temperature increase and the rate of radius contraction drop significantly after the first $\sim 0.2$~Myr. At this point we see that the planet and the core settle in a quasi equilibrium state in which the planet expands very slowly (compared to its dynamical time which is of order 1 yr). The expansion is fuelled by the core growth to about $10\mearth$. 

Now, when the disc is finally removed, stellar radiation starts to heat the planet directly rather than processed through the disc, which increases $T_{\rm irr}$, further slowing the rate of planet cooling. This then leads to the core gaining the upper hand in the heating-cooling balance; the envelope expansion accelerates at that point. This goes on until the envelope is disrupted at $t\approx 1.5$~Myr. 

The disruption releases $2\mj$ of fresh gas and about $70\mearth$ of dust. Note that dust particles outside the core, even as large as $\sim 1$~cm, are well coupled to the gas in the planet (since the planet pre-disruption gas density is much higher than that of the protoplanetary disc), and are therefore disrupted together with it \citep{ChaNayakshin11a,NayakshinCha12}. The secondary protoplanetary disc is born. The initial accretion rate onto the star is quite high, $\dot M_*\sim 2 \times 10^{-8}\msun$~yr$^{-1}$, but subsizes quickly, dropping to below $10^{-10} \msun$~yr$^{-1}$ within $\sim 0.3$~Myr. As with the primary disc, the outer disc lasts somewhat longer  until it is completely removed by photo-evaporation at $t\approx 2$~Myr.

\subsection{Observational Implications}\label{sec:obs}

There are several distinctive observable characteristics of the secondary disc scenario proposed here. First of all, the gas mass of the secondary discs is much smaller than expected on the basis of the two usually available diagnostics. The Steady-State scenario (\S \ref{sec:SS}) would interpret a $t_* = 1.5$~Myr old star accreting at the rate of $\sim 10^{-8}\msun$~yr$^{-1}$ as having a disc with gas mass $M_{\rm d} \sim 2 \dot M_* t_* = 30\mj$, whereas the actual disc mass is less than $2\mj$. Additionally, the planet metallicity at disruption is $Z \approx 8 Z_\odot$, which is much higher than the disc initial pebble abundance of $2 Z_\odot$. Thus one would over-estimate the gas mass by a factor of $\sim$ four if using the `canonical' gas-to-dust ratio (50 for the initial dust abundance in our calculation). This prediction of the theory can be checked in the future by HD molecular line observations. The CO line observations are probably less trustworthy in this case since the luminosity of the disc in this molecule depends not only on the difficult to model disc chemistry but should also be generally proportional to the unknown dust-to-gas ratio.

The most direct observational test of the secondary planet disruption disc scenario is detection or otherwise of the pre-disruption dusty molecular clumps. These may be detectable with  ALMA long-look observations. To show that, the bottom right panel of fig. \ref{fig:TD-Secondary-Disc} plots with the blue dotted line the optically thin dust mass estimate for the clump. This calculation assumes that a clump is imaged but unresolved by ALMA, which may be the most likely case since ALMA beam size for protoplanetary discs tends to from a few AU to 10 AU. Assuming that ALMA detected a source within this beam size, and that an optically thin dust emissivity is used (as usual), the deduced mass is
\begin{equation}
    M_{\rm opt-thin} = \frac{\pi R_{\rm p}^2}{\kappa_\nu} = 
    4.2 \; M_{\rm Moon} \; \left(\frac{R_{\rm p}}{\hbox{1 AU}}\right)^2 \frac{2.3}{\kappa_\nu}\;,
    \label{m-opt-thin}
\end{equation}{}
where the dust opacity $\kappa_\nu$ is scaled to the widely used value of 2.3 cm$^2$/g (for frequency of 230 GHz). Of course the correct dust mass in the clump is much larger, and is given by the solid red curve in the figure, but until the clump is disrupted its emission is consistent with a Lunar to a fraction of an Earth mass of dust. Nevertheless, the high angular resolution of ALMA is likely to be sufficient to discover these clumps if they are there. Paradoxically, however, one may want to point ALMA towards protostars with less massive dust discs which so far have not been systematically targeted by ALMA.

Pre-collapse clumps of TD may be also sufficiently bright in the optical to be detected with high resolution observations (we plan to evaluate this in greater detail in a forthcoming paper). In contrast to the Core Accretion giant planets, which are physically much smaller, TD planets are not expected to emit in the UV even if they are accreting gas because the corresponding shock temperature is in hundreds of Kelvin rather than $\gtrsim 10^4$~K.

\section{Discussion}\label{sec:discussion}

\subsection{Main results}

It was shown in \S \ref{sec:SS} that within the standard interpretation of disc evolution, ALMA planet candidates present a paradox of youth. The inferred gas disc masses in this scenario are large and so the planets must migrate into the sub ten AU region extremely rapidly (figs. \ref{fig:SS} and \ref{fig:CA-EVOL}). To have a decent chance of observing the planets we must require that each of these discs produce tens to hundreds of gas giant planets over its lifetime. This scenario should be rejected based on the observed rarity of giant planets and the available solids mass budget\footnote{In the Appendix we showed that the same conclusion is reached if the gas disc mass is estimated by multiplying the observed dust mass by 100.}.

{\em The paradox of youth plagues any planet formation scenario} as long as the standard disc evolution picture is assumed. The origin of the paradox is in the obtained too high gas disc masses, $M_{\rm d}$, and the assumption that the discs were there (and would be even more massive) at times $t \le t_*$, the age of the system.  In the Appendices we considered possible uncertainties in our results and concluded that they are unlikely to remedy the paradox.


An alternative scenario in which the discs are the result of a deposition of `fresh' gas and dust into the system was considered in \S \ref{sec:TD}. In this case the disc lasts for the viscous time, $t_{\rm visc}$, which depends on the unknown viscosity parameter $\alpha_{\rm v}$. If $t_{\rm visc} \ll t_*$, then several problems are alleviated at once. The inferred disc masses are lower, so the migration time $t_{\rm mig}$ increases, giving us a more reasonable chance to observe the planets; further, we only need $t_{\rm mig}$ to be longer than the viscous time rather than the age of the system. A recent injection of new mass also explains how "old" (e.g., up to $t_* \sim 10$~Myr) protostars can be accreting gas at unexpectedly high rates. In this scenario, the planets may be as old as their stars but the discs are young, and will live for only a fraction of a Million years. 

By considering the planet migration and the dust drift constraints together we arrived at a requirement that the disc viscosity parameter $\alpha_{\rm v}$ needs to be high (fig. \ref{fig:TD-migr} and \S \ref{sec:TD-vs-ALMA}), $\sim 0.1$. This high value of $\alpha_{\rm v}$ is surprising. However, more detailed time-dependent viscous disc calculations show that lower values of $\alpha\sim (1-3)\times 10^{-2}$ may work as well. In our order of magnitude calculation in \S \ref{sec:TD-connection} we estimated the duration during which the secondary disc is observable (before being drained onto the star) as one viscous time (eqs. \ref{tvisc0} and \ref{tvisc100}). The more detailed calculation in \S \ref{sec:example} shows that the star continues to accrete gas for about three viscous times whereas the dusty disc should be observable for almost five viscous times. In a paper in prep. we find that for the very well resolved disc of TW Hydra a value of $\alpha \sim 3\times 10^{-2}$ provides a reasonable fit to the data\footnote{Much smaller values of $\alpha$ are ruled out due to the observed relatively large gas accretion rate in the system.}.

Similarly large values of $\alpha$ were also proposed by other workers, e.g., \cite{ClarkeEtal18}. Finally, $\alpha\gtrsim 10^{-2}$ may not be unreasonable in the view of the recent realisation that large scale magnetic field driven outflows may be very efficient in removing angular momentum of discs \citep{BaiStone13,Bai16}. The corresponding mass of ALMA gas discs is of order the Jovian mass (the left bottom panel of fig. \ref{fig:TD-migr}). Such low mass gas discs have the disc surface density of about 1 g~cm$^{-2}$ on the spatial scales of ALMA discs. They hence may be ionised sufficiently well throughout to maintain a strong coupling to magnetic fields. Further, it is important to stress that the viscosity parameter $\alpha_{\rm v}$ may not necessarily reflect the turbulent viscosity parameter $\alpha_{\rm turb}$ that is often introduced separately and is believed to be much smaller than $0.1$ based on the observations of dust dynamics and line profiles of protoplanetary discs \citep{BroganEtal15,FlahertyEtal18}.

\subsection{The origin of secondary discs}

Given that these low disc masses are consistent with the gas giant planet masses, it is natural to associate these secondary discs with tidal disruptions of one or more dust-rich protoplanets in the context of the Tidal Downsizing theory \citep{Nayakshin_Review}. Such far-out core-feedback initiated disruptions were investigated by \cite{NayakshinCha12}. \cite{Nayakshin16a} explored this scenario further and suggested that HL Tau suspected planets may be formed in such disruptions. In \S \ref{sec:example} we used the code of \cite{Nayakshin16a} to calculate and present an example of a dusty $2\mj$ protoplanet that gets stranded at large separations when the primary disc is dispersed. The planet is over-inflated by a massive solid core growing in its centre, self-destructs, and fields a short lived but bright secondary disc. In this scenario the current sample of bright ALMA discs is dominated by the systems in which protoplanet disruptions occurred very recently. In dimmer systems there are either no wide separation dusty clumps or they are not yet disrupted.

This scenario has a certain observational merit given that dust masses of the ALMA sources used for analysis in this paper lie in the range of $\sim (20-200) \mearth$. This can be surmised from the left panel of fig. \ref{fig:DD} which shows the dust disc mass multiplied by 100. These dust masses fall in the range expected from disruption of a gas giant planet as can be seen from the right bottom panel of fig. \ref{fig:TD-Secondary-Disc}. Additionally, at least for hot Jupiters the inferred metal content of planets more massive than $1\mj$ is between $\sim 30\mearth$ and as much as $\sim 200\mearth$ \citep{ThorngrenEtal15}\footnote{Note that in the TD scenario these planets are physically related to the wide orbit ALMA planet candidates. While hot Jupiters migrated almost all the way to the star,  ALMA planets  were presumably unable to do so.}. These masses are one to two orders of magnitude higher than the logarithmic mean of dust mass of a typical protoplanetary disc. For example, the ODISEA survey \citep{WilliamsEtal19-ODISEA} finds that the logarithmic means of these masses are $\sim 4\mearth$ and $\sim 0.8\mearth$ for class I and class II sources, respectively. In fact \cite{WilliamsEtal19-ODISEA} note that there is a surprising amount of scatter in the dust masses of discs at all ages and they suggest that there may be a `substantial dust regeneration after 1 Myr'. The secondary disc scenario naturally offers a mechanism for such late dust regeneration.

One physical difficulty with the secondary scenario in the TD context -- and possibly the reason why this has not been suggested pre-ALMA as far as a rejuvenation mechanism for protoplanetary discs -- is that the disruption of the planets are required to occur at unexpectedly late times, e.g., $1-10$~Myr. The contraction and collapse time of a uniform composition core-less and isolated gas giant planet can be approximated as $t_{\rm col} \sim 0.5$~Myr $\times (Z/Z_\odot)^{\delta} (M_{\rm p}/\mj)^{-2}$ where $\delta\sim 0.8-1$ \citep[within a factor of two or so as results do depend on the assumed dust opacity;][]{HB11,Nayakshin15c}. This shows that isolated very massive gas giant planets, e.g., $M_{\rm p}\gtrsim 3 \mj$ are unlikely to last Millions of years in the pre-collapse state even if they are very metal rich. Once they collapse via Hydrogen molecule dissociation \citep{Bodenheimer74,Nayakshin15b}, their radius falls to just a few Jupiter radii. This post-collapse state of young planets is known as `hot start' GI planets in the previous literature \citep[e.g.,][]{MarleyEtal07}. Such dense planets cannot be disrupted by their cores or via tides.

External irradiation of planets can delay their contraction and even unbind them \citep[e.g.,][]{CameronEtal82,VazanHelled12}. This effect was indeed taken into account in the model presented in \S \ref{sec:example}. However much more work is warranted as a physically related problem of inflated hot Jupiters shows that additional effects such as Ohmic heating, radiative green house and `mechanical green house', may be important \citep[see, e.g.,][]{GinzburgSari15-inflated-hotJ}. 

Alternatively, secondary discs could be formed by a fresh deposition of gas from outside of the system, as argued by \cite{ManaraEtal18}. The discs in this scenario are "conveyor belts" and are continuously replenished with further matter falling onto the system from the outside. One potential challenge to this scenario is the flat architecture of the ALMA discs that are co-planar with the planets. The orientation of the angular momentum vector of the fresh material is likely to be random \citep[chaotic;][]{BateLodPrin10} and is unlikely to coincide with that of the early gas infall. One therefore could expect the discs to be warped and misaligned with respect to the orbital plane of the planets. However, if the discs are massive enough then planets may readjust to the new orbital plane sufficiently fast.  We leave a detailed evaluation of this scenario for future work.

One potentially promising way of differentiating between the two scenarios for the origin of secondary discs is the gas-to-dust ratio of these discs. If the discs are produced in the disruptions of Tidal Downsizing planets then they must be strongly enriched in metals, and thus dust, compared to the host stars. This is because GI protoplanets are found to accrete pebbles and become pebble-rich \citep{HN18,Baehr19-pebble-accretion}. Furthermore, physical self-consistency of the model requires the protoplanets to be metal rich because Solar metallicity protoplanets are expected to contract much faster because their opacity is lower \citep{HB11}. Protoplanets with a Solar-like composition would not last in their extended disruption-prone state for Millions of years, and their metal content would also be insuffient to make solid cores massive enough to drive their disruption \citep{Nayakshin16a}. Hence TD scenario predicts an enhanced gas-to-dust ratio for the secondary discs. In contrast, fresh material deposited from outside would be of simular composition to the host star, with a gas-to-dust ratio closer to the canonical $\sim 100$.

\subsection{Direct observational tests of the TD secondary disc scenario}

It was argued in \S \ref{sec:obs} that long-look ALMA observations should be able to detect the dusty pre-disruption giant planets in the $\sim$ few Myr old discs if such planets are there. In a paper in preparation (Humphries et al 2020) we find that detecting AU scale optically thick passively irradiated planets (their internal luminosities are quite low) with ALMA in the dust continuum emission is not trivial but not impossible either.  Less bright in the mm-continuum dust emission systems may in fact be more promising targets to look for such planets as they may be easier to disentangle from the protoplanetary disc emission.  Additionally, albedo of dusty molecular gas is $\sim 0.5$ in the optical/NIR wavelengths \citep{WoitkeEtal19}, implying that a significant fraction of the stellar luminosity incident onto the protoplanets could be reflected and hence be observable. Emission from several cold clumps of unknown nature with AU-scale sizes at distance of tens of AU from the host star has been recently detected in PDS 70 \citep{Mesa19-PDS70}, which is one of the objects included in our study.

Additionally, if the disruption process is less abrupt than we assumed here then the clumps may be surrounded by emission elongated mainly along their orbits of both gas and dust that could stand out much better than point-like pre-disruption protoplanets. Our chances of observing this could be bolstered further by a correspondingly longer duration over which such more gradual mass loss could be occurring compared to an abrupt protoplanet disruption. The dynamics of gas, small and large dust is likely to be all different from other extended emission sources such as vortexes. Further work on the process of protoplanet disruption and its observational signatures is well warranted.

\subsection{Connection to Direct Imaging searches for planets}

If such dusty protoplanets are found in significant numbers then this would require a significant re-evaluation of current ideas about the likelyhood of protoplanetary disc fragmentation on planetary mass objects. A number of authors \citep[e.g., see the review by][]{KratterL16} argue that since Direct Imaging observations limit the frequency of occurence of planetary mass objects on wide orbits to only a few \% of FGK stars then GI fragmentation is similarly rare \citep{ViganEtal17}. \cite{HumphriesEtal19} considered one GI gas protoplanet per massive gas disc, and neglected the role of massive cores in the protoplanet evolution. They found that inward migration removes these planets from the wide orbits very efficiently and that most are destroyed by tides within $\sim 10$~AU. They argued that a much larger number of clumps is needed to be fielded at wide separations at $t\sim 0$ to explain both the \cite{ViganEtal17} constraints and the observed `not Core Accretion' population of very massive gas giants inside a few AU \citep{SantosEtal17,Schlaufman18,Adibekyan19,MaldonadoEtal19}. In particular, \cite{HumphriesEtal19} noted that at least 30\% of stars need to host a GI {\em protoplanet} on a wide orbit to explain these observations. 

If the secondary disruption scenario proposed in this paper is correct then the number of dusty gas clumps  born in the disc at $t\approx 0$ must be much larger, e.g., $\sim O(10)$ per star. This could be the case if a typical disc exceeds 100 AU in the embedded phase of star formation and fragments on many protoplanets. To satisfy the stringent Direct Imaging constraints 99\% or more of these clumps needs to be `removed' by either migration into the inner disc,  N-body scatterings with other planets, secondary stars or external stellar mass perturbers, or by late disruptions that are key to the present paper. Whether this is feasible is an open question.

Whatever the outcome of future observational searches for the pre-disruption clumps, the advent of ALMA observations of dusty protoplanetary discs harbouring planet candidates ushers theoretical modelling of planet formation into a new realm. Instead of using very rich but {\em disjoint} sets of constraints on the protoplanetary disc evolution and the statistics of planets in systems {\em no longer forming planets}, which is unfortunately fraught with significant uncertainties,  theorists should now produce joint models of discs and planets evolving in concert to try and address the most modern data.


\section*{Acknowledgements}
The author acknowledges support from STFC grants ST/N000757/1 and ST/M006948/1 to the University of Leicester. The workshop programme `Astrophysical Origins: Pathways from Star Formation to Habitable Planets' held in Erwin Schr\"odinger International Institute in Vienna in the summer of 2019 has been very useful in organising author's thoughts on the subject. The author thanks workshop organisers, Manuel G\"udel, Theresa Lueftinger, Ramon Brasser and Stephen Mojzsis, for hospitality, financial support during his stay in Vienna, and its participants for many lively discussions.



\bibliographystyle{mnras}
\bibliography{nayakshin}

\begin{thebibliography}{}
\makeatletter
\relax
\def\mn@urlcharsother{\let\do\@makeother \do\$\do\&\do\#\do\^\do\_\do\%\do\~}
\def\mn@doi{\begingroup\mn@urlcharsother \@ifnextchar [ {\mn@doi@}
  {\mn@doi@[]}}
\def\mn@doi@[#1]#2{\def\@tempa{#1}\ifx\@tempa\@empty \href
  {http://dx.doi.org/#2} {doi:#2}\else \href {http://dx.doi.org/#2} {#1}\fi
  \endgroup}
\def\mn@eprint#1#2{\mn@eprint@#1:#2::\@nil}
\def\mn@eprint@arXiv#1{\href {http://arxiv.org/abs/#1} {{\tt arXiv:#1}}}
\def\mn@eprint@dblp#1{\href {http://dblp.uni-trier.de/rec/bibtex/#1.xml}
  {dblp:#1}}
\def\mn@eprint@#1:#2:#3:#4\@nil{\def\@tempa {#1}\def\@tempb {#2}\def\@tempc
  {#3}\ifx \@tempc \@empty \let \@tempc \@tempb \let \@tempb \@tempa \fi \ifx
  \@tempb \@empty \def\@tempb {arXiv}\fi \@ifundefined
  {mn@eprint@\@tempb}{\@tempb:\@tempc}{\expandafter \expandafter \csname
  mn@eprint@\@tempb\endcsname \expandafter{\@tempc}}}

\bibitem[\protect\citeauthoryear{{ALMA Partnership} et~al.,}{{ALMA Partnership}
  et~al.}{2015}]{BroganEtal15}
{ALMA Partnership} et~al., 2015, \mn@doi [\apjl] {10.1088/2041-8205/808/1/L3},
  \href {http://adsabs.harvard.edu/abs/2015ApJ...808L...3A} {808, L3}

\bibitem[\protect\citeauthoryear{{Adibekyan}}{{Adibekyan}}{2019}]{Adibekyan19}
{Adibekyan} V.,  2019, arXiv e-prints, \href
  {http://adsabs.harvard.edu/abs/2019arXiv190204493A} {}

\bibitem[\protect\citeauthoryear{{Alexander} \& {Armitage}}{{Alexander} \&
  {Armitage}}{2009}]{AA09}
{Alexander} R.~D.,  {Armitage} P.~J.,  2009, \mn@doi [\apj]
  {10.1088/0004-637X/704/2/989}, \href
  {http://adsabs.harvard.edu/abs/2009ApJ...704..989A} {704, 989}

\bibitem[\protect\citeauthoryear{{Alexander}, {Pascucci}, {Andrews}, {Armitage}
   \& {Cieza}}{{Alexander} et~al.}{2014}]{AlexanderREtal14a}
{Alexander} R.,  {Pascucci} I.,  {Andrews} S.,  {Armitage} P.,   {Cieza} L.,
  2014, \mn@doi [Protostars and Planets VI]
  {10.2458/azu_uapress_9780816531240-ch021}, \href
  {http://adsabs.harvard.edu/abs/2014prpl.conf..475A} {pp 475--496}

\bibitem[\protect\citeauthoryear{{Alibert}, {Mordasini}, {Benz}  \&
  {Winisdoerffer}}{{Alibert} et~al.}{2005}]{AlibertEtal05}
{Alibert} Y.,  {Mordasini} C.,  {Benz} W.,   {Winisdoerffer} C.,  2005, \mn@doi
  [\aap] {10.1051/0004-6361:20042032}, \href
  {http://adsabs.harvard.edu/abs/2005A%26A...434..343A} {434, 343}

\bibitem[\protect\citeauthoryear{{Andrews} et~al.,}{{Andrews}
  et~al.}{2016}]{AndrewsEtal16}
{Andrews} S.~M.,  et~al., 2016, \mn@doi [\apjl] {10.3847/2041-8205/820/2/L40},
  \href {https://ui.adsabs.harvard.edu/abs/2016ApJ...820L..40A} {820, L40}

\bibitem[\protect\citeauthoryear{{Andrews} et~al.,}{{Andrews}
  et~al.}{2018}]{Dsharp1}
{Andrews} S.~M.,  et~al., 2018, \mn@doi [\apjl] {10.3847/2041-8213/aaf741},
  \href {http://adsabs.harvard.edu/abs/2018ApJ...869L..41A} {869, L41}

\bibitem[\protect\citeauthoryear{{Ansdell} et~al.,}{{Ansdell}
  et~al.}{2018}]{AnsdellEtal18}
{Ansdell} M.,  et~al., 2018, \mn@doi [\apj] {10.3847/1538-4357/aab890}, \href
  {https://ui.adsabs.harvard.edu/abs/2018ApJ...859...21A} {859, 21}

\bibitem[\protect\citeauthoryear{{Baehr} \& {Klahr}}{{Baehr} \&
  {Klahr}}{2019}]{Baehr19-pebble-accretion}
{Baehr} H.,  {Klahr} H.,  2019, \mn@doi [\apj] {10.3847/1538-4357/ab2f85},
  \href {https://ui.adsabs.harvard.edu/abs/2019ApJ...881..162B} {881, 162}

\bibitem[\protect\citeauthoryear{{Bai}}{{Bai}}{2016}]{Bai16}
{Bai} X.-N.,  2016, \mn@doi [\apj] {10.3847/0004-637X/821/2/80}, \href
  {https://ui.adsabs.harvard.edu/abs/2016ApJ...821...80B} {821, 80}

\bibitem[\protect\citeauthoryear{{Bai} \& {Stone}}{{Bai} \&
  {Stone}}{2013}]{BaiStone13}
{Bai} X.-N.,  {Stone} J.~M.,  2013, \mn@doi [\apj]
  {10.1088/0004-637X/769/1/76}, \href
  {https://ui.adsabs.harvard.edu/abs/2013ApJ...769...76B} {769, 76}

\bibitem[\protect\citeauthoryear{{Baraffe}, {Vorobyov}  \&
  {Chabrier}}{{Baraffe} et~al.}{2012}]{BaraffeEtal12}
{Baraffe} I.,  {Vorobyov} E.,   {Chabrier} G.,  2012, \mn@doi [\apj]
  {10.1088/0004-637X/756/2/118}, \href
  {http://ukads.nottingham.ac.uk/abs/2012ApJ...756..118B} {756, 118}

\bibitem[\protect\citeauthoryear{{Baruteau}, {Meru}  \&
  {Paardekooper}}{{Baruteau} et~al.}{2011}]{BaruteauEtal11}
{Baruteau} C.,  {Meru} F.,   {Paardekooper} S.-J.,  2011, \mn@doi [\mnras]
  {10.1111/j.1365-2966.2011.19172.x}, \href
  {http://adsabs.harvard.edu/abs/2011MNRAS.416.1971B} {416, 1971}

\bibitem[\protect\citeauthoryear{{Bate}, {Lodato}  \& {Pringle}}{{Bate}
  et~al.}{2010}]{BateLodPrin10}
{Bate} M.~R.,  {Lodato} G.,   {Pringle} J.~E.,  2010, \mn@doi [\mnras]
  {10.1111/j.1365-2966.2009.15773.x}, \href
  {http://ukads.nottingham.ac.uk/abs/2010MNRAS.401.1505B} {401, 1505}

\bibitem[\protect\citeauthoryear{{Birnstiel}, {Dullemond}  \&
  {Brauer}}{{Birnstiel} et~al.}{2009}]{Birnstiel09}
{Birnstiel} T.,  {Dullemond} C.~P.,   {Brauer} F.,  2009, \mn@doi [\aap]
  {10.1051/0004-6361/200912452}, \href
  {https://ui.adsabs.harvard.edu/abs/2009A%26A...503L...5B} {503, L5}

\bibitem[\protect\citeauthoryear{{Birnstiel}, {Klahr}  \&
  {Ercolano}}{{Birnstiel} et~al.}{2012}]{BirnstielEtal12}
{Birnstiel} T.,  {Klahr} H.,   {Ercolano} B.,  2012, \mn@doi [\aap]
  {10.1051/0004-6361/201118136}, \href
  {https://ui.adsabs.harvard.edu/abs/2012A%26A...539A.148B} {539, A148}

\bibitem[\protect\citeauthoryear{{Bitsch}, {Lambrechts}  \&
  {Johansen}}{{Bitsch} et~al.}{2015}]{BitschEtal15}
{Bitsch} B.,  {Lambrechts} M.,   {Johansen} A.,  2015, \mn@doi [\aap]
  {10.1051/0004-6361/201526463}, \href
  {https://ui.adsabs.harvard.edu/abs/2015A%26A...582A.112B} {582, A112}

\bibitem[\protect\citeauthoryear{{Bitsch}, {Izidoro}, {Johansen}, {Raymond},
  {Morbidelli}, {Lambrechts}  \& {Jacobson}}{{Bitsch}
  et~al.}{2019}]{BitschEtal19}
{Bitsch} B.,  {Izidoro} A.,  {Johansen} A.,  {Raymond} S.~N.,  {Morbidelli} A.,
   {Lambrechts} M.,   {Jacobson} S.~A.,  2019, \mn@doi [\aap]
  {10.1051/0004-6361/201834489}, \href
  {https://ui.adsabs.harvard.edu/abs/2019A%26A...623A..88B} {623, A88}

\bibitem[\protect\citeauthoryear{{Bodenheimer}}{{Bodenheimer}}{1974}]{Bodenheimer74}
{Bodenheimer} P.,  1974, \mn@doi [Icarus] {10.1016/0019-1035(74)90050-5}, \href
  {http://ukads.nottingham.ac.uk/abs/1974Icar...23..319B} {23, 319}

\bibitem[\protect\citeauthoryear{{Boley} \& {Durisen}}{{Boley} \&
  {Durisen}}{2010}]{BoleyDurisen10}
{Boley} A.~C.,  {Durisen} R.~H.,  2010, \mn@doi [\apj]
  {10.1088/0004-637X/724/1/618}, \href
  {http://adsabs.harvard.edu/abs/2010ApJ...724..618B} {724, 618}

\bibitem[\protect\citeauthoryear{{Boley}, {Hayfield}, {Mayer}  \&
  {Durisen}}{{Boley} et~al.}{2010}]{BoleyEtal10}
{Boley} A.~C.,  {Hayfield} T.,  {Mayer} L.,   {Durisen} R.~H.,  2010, \mn@doi
  [Icarus] {10.1016/j.icarus.2010.01.015}, \href
  {http://ukads.nottingham.ac.uk/abs/2010Icar..207..509B} {207, 509}

\bibitem[\protect\citeauthoryear{{Boley}, {Helled}  \& {Payne}}{{Boley}
  et~al.}{2011}]{BoleyEtal11a}
{Boley} A.~C.,  {Helled} R.,   {Payne} M.~J.,  2011, \mn@doi [\apj]
  {10.1088/0004-637X/735/1/30}, \href
  {http://adsabs.harvard.edu/abs/2011ApJ...735...30B} {735, 30}

\bibitem[\protect\citeauthoryear{{Boss}}{{Boss}}{1998}]{Boss98}
{Boss} A.~P.,  1998, \mn@doi [\apj] {10.1086/306036}, \href
  {http://adsabs.harvard.edu/abs/1998ApJ...503..923B} {503, 923}

\bibitem[\protect\citeauthoryear{{Bridges}, {Changela}, {Nayakshin}, {Starkey}
  \& {Franchi}}{{Bridges} et~al.}{2012}]{BridgesEtal12a}
{Bridges} J.~C.,  {Changela} H.~G.,  {Nayakshin} S.,  {Starkey} N.~A.,
  {Franchi} I.~A.,  2012, \mn@doi [Earth and Planetary Science Letters]
  {10.1016/j.epsl.2012.06.011}, \href
  {http://adsabs.harvard.edu/abs/2012E%26PSL.341..186B} {341, 186}

\bibitem[\protect\citeauthoryear{{Cameron}, {Decampli}  \&
  {Bodenheimer}}{{Cameron} et~al.}{1982}]{CameronEtal82}
{Cameron} A.~G.~W.,  {Decampli} W.~M.,   {Bodenheimer} P.,  1982, \mn@doi
  [Icarus] {10.1016/0019-1035(82)90038-0}, \href
  {http://ukads.nottingham.ac.uk/abs/1982Icar...49..298C} {49, 298}

\bibitem[\protect\citeauthoryear{{Cha} \& {Nayakshin}}{{Cha} \&
  {Nayakshin}}{2011}]{ChaNayakshin11a}
{Cha} S.-H.,  {Nayakshin} S.,  2011, \mn@doi [\mnras]
  {10.1111/j.1365-2966.2011.18953.x}, \href
  {http://adsabs.harvard.edu/abs/2011MNRAS.415.3319C} {415, 3319}

\bibitem[\protect\citeauthoryear{{Clarke} et~al.,}{{Clarke}
  et~al.}{2018}]{ClarkeEtal18}
{Clarke} C.~J.,  et~al., 2018, \mn@doi [\apjl] {10.3847/2041-8213/aae36b},
  \href {http://adsabs.harvard.edu/abs/2018ApJ...866L...6C} {866, L6}

\bibitem[\protect\citeauthoryear{{Crida}, {Morbidelli}  \& {Masset}}{{Crida}
  et~al.}{2006}]{CridaEtal06}
{Crida} A.,  {Morbidelli} A.,   {Masset} F.,  2006, \mn@doi [\icarus]
  {10.1016/j.icarus.2005.10.007}, \href
  {http://adsabs.harvard.edu/abs/2006Icar..181..587C} {181, 587}

\bibitem[\protect\citeauthoryear{{Dipierro}, {Price}, {Laibe}, {Hirsh},
  {Cerioli}  \& {Lodato}}{{Dipierro} et~al.}{2015}]{DipierroEtal15}
{Dipierro} G.,  {Price} D.,  {Laibe} G.,  {Hirsh} K.,  {Cerioli} A.,   {Lodato}
  G.,  2015, \mn@doi [\mnras] {10.1093/mnrasl/slv105}, \href
  {http://adsabs.harvard.edu/abs/2015MNRAS.453L..73D} {453, L73}

\bibitem[\protect\citeauthoryear{{Dipierro}, {Laibe}, {Price}  \&
  {Lodato}}{{Dipierro} et~al.}{2016}]{DipierroEtal16a}
{Dipierro} G.,  {Laibe} G.,  {Price} D.~J.,   {Lodato} G.,  2016, \mn@doi
  [\mnras] {10.1093/mnrasl/slw032}, \href
  {http://adsabs.harvard.edu/abs/2016MNRAS.tmpL..16D} {}

\bibitem[\protect\citeauthoryear{{Dipierro} et~al.,}{{Dipierro}
  et~al.}{2018}]{DipierroEtal18}
{Dipierro} G.,  et~al., 2018, \mn@doi [\mnras] {10.1093/mnras/sty181}, \href
  {https://ui.adsabs.harvard.edu/abs/2018MNRAS.475.5296D} {475, 5296}

\bibitem[\protect\citeauthoryear{{Dullemond} \& {Dominik}}{{Dullemond} \&
  {Dominik}}{2005}]{DD05}
{Dullemond} C.~P.,  {Dominik} C.,  2005, \mn@doi [\aap]
  {10.1051/0004-6361:20042080}, \href
  {http://ukads.nottingham.ac.uk/abs/2005A%26A...434..971D} {434, 971}

\bibitem[\protect\citeauthoryear{{Dullemond} et~al.,}{{Dullemond}
  et~al.}{2018}]{DSHARP-6}
{Dullemond} C.~P.,  et~al., 2018, \mn@doi [\apjl] {10.3847/2041-8213/aaf742},
  \href {http://adsabs.harvard.edu/abs/2018ApJ...869L..46D} {869, L46}

\bibitem[\protect\citeauthoryear{{Dunham} \& {Vorobyov}}{{Dunham} \&
  {Vorobyov}}{2012}]{DunhamVorobyov12}
{Dunham} M.~M.,  {Vorobyov} E.~I.,  2012, \mn@doi [\apj]
  {10.1088/0004-637X/747/1/52}, \href
  {http://adsabs.harvard.edu/abs/2012ApJ...747...52D} {747, 52}

\bibitem[\protect\citeauthoryear{{Fabrycky} et~al.,}{{Fabrycky}
  et~al.}{2014}]{FabryckyEtal14}
{Fabrycky} D.~C.,  et~al., 2014, \mn@doi [\apj] {10.1088/0004-637X/790/2/146},
  \href {http://adsabs.harvard.edu/abs/2014ApJ...790..146F} {790, 146}

\bibitem[\protect\citeauthoryear{{Fernandes}, {Mulders}, {Pascucci},
  {Mordasini}  \& {Emsenhuber}}{{Fernandes} et~al.}{2019}]{FernandesEtal19}
{Fernandes} R.~B.,  {Mulders} G.~D.,  {Pascucci} I.,  {Mordasini} C.,
  {Emsenhuber} A.,  2019, \mn@doi [\apj] {10.3847/1538-4357/ab0300}, \href
  {https://ui.adsabs.harvard.edu/abs/2019ApJ...874...81F} {874, 81}

\bibitem[\protect\citeauthoryear{{Flagg}, {Johns-Krull}, {Nofi}, {Llama},
  {Prato}, {Sullivan}, {Jaffe}  \& {Mace}}{{Flagg} et~al.}{2019}]{FlaggEtal19}
{Flagg} L.,  {Johns-Krull} C.~M.,  {Nofi} L.,  {Llama} J.,  {Prato} L.,
  {Sullivan} K.,  {Jaffe} D.~T.,   {Mace} G.,  2019, \mn@doi [\apjl]
  {10.3847/2041-8213/ab276d}, \href
  {https://ui.adsabs.harvard.edu/abs/2019ApJ...878L..37F} {878, L37}

\bibitem[\protect\citeauthoryear{{Flaherty}, {Hughes}, {Teague}, {Simon},
  {Andrews}  \& {Wilner}}{{Flaherty} et~al.}{2018}]{FlahertyEtal18}
{Flaherty} K.~M.,  {Hughes} A.~M.,  {Teague} R.,  {Simon} J.~B.,  {Andrews}
  S.~M.,   {Wilner} D.~J.,  2018, \mn@doi [\apj] {10.3847/1538-4357/aab615},
  \href {http://adsabs.harvard.edu/abs/2018ApJ...856..117F} {856, 117}

\bibitem[\protect\citeauthoryear{{Fletcher}, {Nayakshin}, {Stamatellos},
  {Dehnen}, {Meru}, {Mayer}, {Deng}  \& {Rice}}{{Fletcher}
  et~al.}{2019}]{FletcherEtal19}
{Fletcher} M.,  {Nayakshin} S.,  {Stamatellos} D.,  {Dehnen} W.,  {Meru} F.,
  {Mayer} L.,  {Deng} H.,   {Rice} K.,  2019, \mn@doi [\mnras]
  {10.1093/mnras/stz1123}, \href
  {https://ui.adsabs.harvard.edu/abs/2019MNRAS.486.4398F} {486, 4398}

\bibitem[\protect\citeauthoryear{{Forgan}}{{Forgan}}{2019}]{Forgan19}
{Forgan} D.~H.,  2019, \mn@doi [\mnras] {10.1093/mnras/stz494}, \href
  {https://ui.adsabs.harvard.edu/abs/2019MNRAS.485.4465F} {485, 4465}

\bibitem[\protect\citeauthoryear{{Forgan} \& {Rice}}{{Forgan} \&
  {Rice}}{2011}]{ForganRice11}
{Forgan} D.,  {Rice} K.,  2011, \mn@doi [\mnras]
  {10.1111/j.1365-2966.2011.19380.x}, \href
  {http://adsabs.harvard.edu/abs/2011MNRAS.417.1928F} {417, 1928}

\bibitem[\protect\citeauthoryear{{Forgan} \& {Rice}}{{Forgan} \&
  {Rice}}{2013}]{ForganRice13b}
{Forgan} D.,  {Rice} K.,  2013, \mn@doi [\mnras] {10.1093/mnras/stt672}, \href
  {http://adsabs.harvard.edu/abs/2013MNRAS.432.3168F} {432, 3168}

\bibitem[\protect\citeauthoryear{{Galvagni} \& {Mayer}}{{Galvagni} \&
  {Mayer}}{2014}]{GalvagniMayer14}
{Galvagni} M.,  {Mayer} L.,  2014, \mn@doi [\mnras] {10.1093/mnras/stt2108},
  \href {http://adsabs.harvard.edu/abs/2014MNRAS.437.2909G} {437, 2909}

\bibitem[\protect\citeauthoryear{{Gammie}}{{Gammie}}{2001}]{Gammie01}
{Gammie} C.~F.,  2001, \apj, \href
  {http://cdsads.u-strasbg.fr/cgi-bin/nph-bib_query?bibcode=2001ApJ...553..174G&amp;db_key=AST}
  {553, 174}

\bibitem[\protect\citeauthoryear{{Ginzburg} \& {Sari}}{{Ginzburg} \&
  {Sari}}{2015}]{GinzburgSari15-inflated-hotJ}
{Ginzburg} S.,  {Sari} R.,  2015, \mn@doi [\apj] {10.1088/0004-637X/803/2/111},
  \href {https://ui.adsabs.harvard.edu/abs/2015ApJ...803..111G} {803, 111}

\bibitem[\protect\citeauthoryear{{Haisch}, {Lada}  \& {Lada}}{{Haisch}
  et~al.}{2001}]{HaischEtal01}
{Haisch} Jr. K.~E.,  {Lada} E.~A.,   {Lada} C.~J.,  2001, \mn@doi [\apjl]
  {10.1086/320685}, \href {http://adsabs.harvard.edu/abs/2001ApJ...553L.153H}
  {553, L153}

\bibitem[\protect\citeauthoryear{{Hall}, {Forgan}  \& {Rice}}{{Hall}
  et~al.}{2017}]{HallCEtal17}
{Hall} C.,  {Forgan} D.,   {Rice} K.,  2017, \mn@doi [\mnras]
  {10.1093/mnras/stx1244}, \href
  {http://adsabs.harvard.edu/abs/2017MNRAS.470.2517H} {470, 2517}

\bibitem[\protect\citeauthoryear{{Hartmann} \& {Kenyon}}{{Hartmann} \&
  {Kenyon}}{1996}]{HK96}
{Hartmann} L.,  {Kenyon} S.~J.,  1996, \mn@doi [\araa]
  {10.1146/annurev.astro.34.1.207}, \href
  {http://adsabs.harvard.edu/abs/1996ARA%26A..34..207H} {34, 207}

\bibitem[\protect\citeauthoryear{{Helled} \& {Bodenheimer}}{{Helled} \&
  {Bodenheimer}}{2011}]{HB11}
{Helled} R.,  {Bodenheimer} P.,  2011, \mn@doi [\icarus]
  {10.1016/j.icarus.2010.09.024}, \href
  {http://adsabs.harvard.edu/abs/2011Icar..211..939H} {211, 939}

\bibitem[\protect\citeauthoryear{{Helled}, {Podolak}  \& {Kovetz}}{{Helled}
  et~al.}{2006}]{HelledEtal06}
{Helled} R.,  {Podolak} M.,   {Kovetz} A.,  2006, \mn@doi [\icarus]
  {10.1016/j.icarus.2006.06.011}, \href
  {http://adsabs.harvard.edu/abs/2006Icar..185...64H} {185, 64}

\bibitem[\protect\citeauthoryear{{Helled}, {Podolak}  \& {Kovetz}}{{Helled}
  et~al.}{2008}]{HelledEtal08}
{Helled} R.,  {Podolak} M.,   {Kovetz} A.,  2008, \mn@doi [Icarus]
  {10.1016/j.icarus.2008.01.007}, \href
  {http://ukads.nottingham.ac.uk/abs/2008Icar..195..863H} {195, 863}

\bibitem[\protect\citeauthoryear{{Hendler} et~al.,}{{Hendler}
  et~al.}{2018}]{HendlerEtal18}
{Hendler} N.~P.,  et~al., 2018, \mn@doi [\mnras] {10.1093/mnrasl/slx184}, \href
  {https://ui.adsabs.harvard.edu/abs/2018MNRAS.475L..62H} {475, L62}

\bibitem[\protect\citeauthoryear{{Huang} et~al.,}{{Huang}
  et~al.}{2018}]{Dsharp2}
{Huang} J.,  et~al., 2018, \mn@doi [\apjl] {10.3847/2041-8213/aaf740}, \href
  {http://adsabs.harvard.edu/abs/2018ApJ...869L..42H} {869, L42}

\bibitem[\protect\citeauthoryear{{Humphries} \& {Nayakshin}}{{Humphries} \&
  {Nayakshin}}{2018}]{HN18}
{Humphries} R.~J.,  {Nayakshin} S.,  2018, \mn@doi [\mnras]
  {10.1093/mnras/sty569}, \href
  {http://ukads.nottingham.ac.uk/abs/2018MNRAS.tmp..595H} {}

\bibitem[\protect\citeauthoryear{{Humphries}, {Vazan}, {Bonavita}, {Helled}  \&
  {Nayakshin}}{{Humphries} et~al.}{2019}]{HumphriesEtal19}
{Humphries} J.,  {Vazan} A.,  {Bonavita} M.,  {Helled} R.,   {Nayakshin} S.,
  2019, \mn@doi [\mnras] {10.1093/mnras/stz2006}, \href
  {https://ui.adsabs.harvard.edu/abs/2019MNRAS.tmp.1947H} {p.~1947}

\bibitem[\protect\citeauthoryear{{Ida} \& {Lin}}{{Ida} \&
  {Lin}}{2004}]{IdaLin04a}
{Ida} S.,  {Lin} D.~N.~C.,  2004, \mn@doi [\apj] {10.1086/381724}, \href
  {http://adsabs.harvard.edu/abs/2004ApJ...604..388I} {604, 388}

\bibitem[\protect\citeauthoryear{{Ikoma}, {Nakazawa}  \& {Emori}}{{Ikoma}
  et~al.}{2000}]{IkomaEtal00}
{Ikoma} M.,  {Nakazawa} K.,   {Emori} H.,  2000, \mn@doi [\apj]
  {10.1086/309050}, \href
  {http://ukads.nottingham.ac.uk/abs/2000ApJ...537.1013I} {537, 1013}

\bibitem[\protect\citeauthoryear{{Ilee} et~al.,}{{Ilee}
  et~al.}{2017}]{IleeEtal17}
{Ilee} J.~D.,  et~al., 2017, \mn@doi [\mnras] {10.1093/mnras/stx1966}, \href
  {http://adsabs.harvard.edu/abs/2017MNRAS.472..189I} {472, 189}

\bibitem[\protect\citeauthoryear{{Ivanov}, {Papaloizou}  \&
  {Polnarev}}{{Ivanov} et~al.}{1999}]{IvanovEtal99}
{Ivanov} P.~B.,  {Papaloizou} J.~C.~B.,   {Polnarev} A.~G.,  1999, \mnras,
  \href {http://adsabs.harvard.edu/abs/1999MNRAS.307...79I} {307, 79}

\bibitem[\protect\citeauthoryear{{Johansen} \& {Lacerda}}{{Johansen} \&
  {Lacerda}}{2010}]{JohansenLacerda10}
{Johansen} A.,  {Lacerda} P.,  2010, \mn@doi [\mnras]
  {10.1111/j.1365-2966.2010.16309.x}, \href
  {http://adsabs.harvard.edu/abs/2010MNRAS.404..475J} {404, 475}

\bibitem[\protect\citeauthoryear{{Jones}, {Pringle}  \& {Alexander}}{{Jones}
  et~al.}{2012}]{JPA12}
{Jones} M.~G.,  {Pringle} J.~E.,   {Alexander} R.~D.,  2012, \mn@doi [\mnras]
  {10.1111/j.1365-2966.2011.19730.x}, \href
  {http://adsabs.harvard.edu/abs/2012MNRAS.419..925J} {419, 925}

\bibitem[\protect\citeauthoryear{{Kenyon} \& {Luu}}{{Kenyon} \&
  {Luu}}{1999}]{KL99}
{Kenyon} S.~J.,  {Luu} J.~X.,  1999, \mn@doi [\aj] {10.1086/300969}, \href
  {http://adsabs.harvard.edu/abs/1999AJ....118.1101K} {118, 1101}

\bibitem[\protect\citeauthoryear{{Kratter} \& {Lodato}}{{Kratter} \&
  {Lodato}}{2016}]{KratterL16}
{Kratter} K.~M.,  {Lodato} G.,  2016, preprint, \href
  {http://adsabs.harvard.edu/abs/2016arXiv160301280K} {} (\mn@eprint {arXiv}
  {1603.01280})

\bibitem[\protect\citeauthoryear{{Kratter}, {Murray-Clay}  \&
  {Youdin}}{{Kratter} et~al.}{2010}]{KratterEtal10}
{Kratter} K.~M.,  {Murray-Clay} R.~A.,   {Youdin} A.~N.,  2010, \mn@doi [\apj]
  {10.1088/0004-637X/710/2/1375}, \href
  {http://adsabs.harvard.edu/abs/2010ApJ...710.1375K} {710, 1375}

\bibitem[\protect\citeauthoryear{{Kuiper}}{{Kuiper}}{1951}]{Kuiper51b}
{Kuiper} G.~P.,  1951, \mn@doi [Proceedings of the National Academy of Science]
  {10.1073/pnas.37.1.1}, \href
  {http://adsabs.harvard.edu/abs/1951PNAS...37....1K} {37, 1}

\bibitem[\protect\citeauthoryear{{Lambrechts} \& {Johansen}}{{Lambrechts} \&
  {Johansen}}{2012}]{LambrechtsJ12}
{Lambrechts} M.,  {Johansen} A.,  2012, \mn@doi [\aap]
  {10.1051/0004-6361/201219127}, \href
  {http://adsabs.harvard.edu/abs/2012A%26A...544A..32L} {544, A32}

\bibitem[\protect\citeauthoryear{{Lambrechts}, {Johansen}  \&
  {Morbidelli}}{{Lambrechts} et~al.}{2014}]{LambrechtsEtal14}
{Lambrechts} M.,  {Johansen} A.,   {Morbidelli} A.,  2014, \mn@doi [\aap]
  {10.1051/0004-6361/201423814}, \href
  {http://adsabs.harvard.edu/abs/2014A%26A...572A..35L} {572, A35}

\bibitem[\protect\citeauthoryear{{Liu} et~al.,}{{Liu} et~al.}{2019}]{LiuEtal19}
{Liu} Y.,  et~al., 2019, \mn@doi [\aap] {10.1051/0004-6361/201834157}, \href
  {https://ui.adsabs.harvard.edu/abs/2019A%26A...622A..75L} {622, A75}

\bibitem[\protect\citeauthoryear{{Lodato} \& {Rice}}{{Lodato} \&
  {Rice}}{2005}]{LodatoRice05}
{Lodato} G.,  {Rice} W.~K.~M.,  2005, \mn@doi [\mnras]
  {10.1111/j.1365-2966.2005.08875.x}, \href
  {http://adsabs.harvard.edu/abs/2005MNRAS.358.1489L} {358, 1489}

\bibitem[\protect\citeauthoryear{{Lodato} et~al.,}{{Lodato}
  et~al.}{2019}]{LodatoEtal19}
{Lodato} G.,  et~al., 2019, \mn@doi [\mnras] {10.1093/mnras/stz913}, \href
  {http://adsabs.harvard.edu/abs/2019MNRAS.486..453L} {486, 453}

\bibitem[\protect\citeauthoryear{{Long} et~al.,}{{Long}
  et~al.}{2018}]{LongEtal18}
{Long} F.,  et~al., 2018, preprint, \href
  {http://adsabs.harvard.edu/abs/2018arXiv181006044L} {} (\mn@eprint {arXiv}
  {1810.06044})

\bibitem[\protect\citeauthoryear{{Machida}, {Inutsuka}  \&
  {Matsumoto}}{{Machida} et~al.}{2010}]{MachidaEtal10}
{Machida} M.~N.,  {Inutsuka} S.,   {Matsumoto} T.,  2010, \mn@doi [\apj]
  {10.1088/0004-637X/724/2/1006}, \href
  {http://adsabs.harvard.edu/abs/2010ApJ...724.1006M} {724, 1006}

\bibitem[\protect\citeauthoryear{{Macias} et~al.,}{{Macias}
  et~al.}{2019}]{MaciasEtal19}
{Macias} E.,  et~al., 2019, arXiv e-prints, \href
  {https://ui.adsabs.harvard.edu/abs/2019arXiv190707277M} {}

\bibitem[\protect\citeauthoryear{{Maldonado}, {Villaver}, {Eiroa}  \&
  {Micela}}{{Maldonado} et~al.}{2019}]{MaldonadoEtal19}
{Maldonado} J.,  {Villaver} E.,  {Eiroa} C.,   {Micela} G.,  2019, arXiv
  e-prints, \href {http://adsabs.harvard.edu/abs/2019arXiv190301141M} {}

\bibitem[\protect\citeauthoryear{{Manara}, {Morbidelli}  \& {Guillot}}{{Manara}
  et~al.}{2018}]{ManaraEtal18}
{Manara} C.~F.,  {Morbidelli} A.,   {Guillot} T.,  2018, preprint, \href
  {http://adsabs.harvard.edu/abs/2018arXiv180907374M} {} (\mn@eprint {arXiv}
  {1809.07374})

\bibitem[\protect\citeauthoryear{{Manara}, {Mordasini}, {Testi}, {Williams},
  {Miotello}, {Lodato}  \& {Emsenhuber}}{{Manara}
  et~al.}{2019}]{Manara19-ALMA-va-PopSyn}
{Manara} C.~F.,  {Mordasini} C.,  {Testi} L.,  {Williams} J.~P.,  {Miotello}
  A.,  {Lodato} G.,   {Emsenhuber} A.,  2019, arXiv e-prints, \href
  {https://ui.adsabs.harvard.edu/abs/2019arXiv190908485M} {p. arXiv:1909.08485}

\bibitem[\protect\citeauthoryear{{Marley}, {Fortney}, {Hubickyj}, {Bodenheimer}
   \& {Lissauer}}{{Marley} et~al.}{2007}]{MarleyEtal07}
{Marley} M.~S.,  {Fortney} J.~J.,  {Hubickyj} O.,  {Bodenheimer} P.,
  {Lissauer} J.~J.,  2007, \mn@doi [\apj] {10.1086/509759}, \href
  {http://adsabs.harvard.edu/abs/2007ApJ...655..541M} {655, 541}

\bibitem[\protect\citeauthoryear{{Marois}, {Zuckerman}, {Konopacky},
  {Macintosh}  \& {Barman}}{{Marois} et~al.}{2010}]{MaroisEtal10}
{Marois} C.,  {Zuckerman} B.,  {Konopacky} Q.~M.,  {Macintosh} B.,   {Barman}
  T.,  2010, \mn@doi [\nat] {10.1038/nature09684}, \href
  {http://adsabs.harvard.edu/abs/2010Natur.468.1080M} {468, 1080}

\bibitem[\protect\citeauthoryear{{McCrea} \& {Williams}}{{McCrea} \&
  {Williams}}{1965}]{McCreaWilliams65}
{McCrea} W.~H.,  {Williams} I.~P.,  1965, Royal Society of London Proceedings
  Series A, \href {http://adsabs.harvard.edu/abs/1965RSPSA.287..143M} {287,
  143}

\bibitem[\protect\citeauthoryear{{Mesa} et~al.,}{{Mesa}
  et~al.}{2019}]{Mesa19-PDS70}
{Mesa} D.,  et~al., 2019, \mn@doi [\aap] {10.1051/0004-6361/201936764}, \href
  {https://ui.adsabs.harvard.edu/abs/2019A&A...632A..25M} {632, A25}

\bibitem[\protect\citeauthoryear{{Michael}, {Durisen}  \& {Boley}}{{Michael}
  et~al.}{2011}]{MichaelEtal11}
{Michael} S.,  {Durisen} R.~H.,   {Boley} A.~C.,  2011, \mn@doi [\apjl]
  {10.1088/2041-8205/737/2/L42}, \href
  {http://adsabs.harvard.edu/abs/2011ApJ...737L..42M} {737, L42+}

\bibitem[\protect\citeauthoryear{{Mordasini}}{{Mordasini}}{2018}]{Mordasini18}
{Mordasini} C.,  2018, {Planetary Population Synthesis}.
p.~143, \mn@doi{10.1007/978-3-319-55333-7_143}

\bibitem[\protect\citeauthoryear{{Mordasini}, {Alibert}, {Benz}  \&
  {Naef}}{{Mordasini} et~al.}{2009}]{MordasiniEtal09b}
{Mordasini} C.,  {Alibert} Y.,  {Benz} W.,   {Naef} D.,  2009, \mn@doi [\aap]
  {10.1051/0004-6361/200810697}, \href
  {http://adsabs.harvard.edu/abs/2009A%26A...501.1161M} {501, 1161}

\bibitem[\protect\citeauthoryear{{Mordasini}, {Alibert}, {Klahr}  \&
  {Henning}}{{Mordasini} et~al.}{2012}]{MordasiniEtal12a}
{Mordasini} C.,  {Alibert} Y.,  {Klahr} H.,   {Henning} T.,  2012, \mn@doi
  [\aap] {10.1051/0004-6361/201118457}, \href
  {http://adsabs.harvard.edu/abs/2012A%26A...547A.111M} {547, A111}

\bibitem[\protect\citeauthoryear{{M{\"u}ller}, {Helled}  \&
  {Mayer}}{{M{\"u}ller} et~al.}{2018}]{MullerEtal18}
{M{\"u}ller} S.,  {Helled} R.,   {Mayer} L.,  2018, \mn@doi [\apj]
  {10.3847/1538-4357/aaa840}, \href
  {http://ukads.nottingham.ac.uk/abs/2018ApJ...854..112M} {854, 112}

\bibitem[\protect\citeauthoryear{{Nayakshin}}{{Nayakshin}}{2010a}]{Nayakshin10c}
{Nayakshin} S.,  2010a, \mn@doi [\mnras] {10.1111/j.1745-3933.2010.00923.x},
  \href {http://adsabs.harvard.edu/abs/2010MNRAS.408L..36N} {408, L36}

\bibitem[\protect\citeauthoryear{{Nayakshin}}{{Nayakshin}}{2010b}]{Nayakshin10a}
{Nayakshin} S.,  2010b, \mn@doi [\mnras] {10.1111/j.1365-2966.2010.17289.x},
  \href {http://adsabs.harvard.edu/abs/2010MNRAS.408.2381N} {408, 2381}

\bibitem[\protect\citeauthoryear{{Nayakshin}}{{Nayakshin}}{2015a}]{Nayakshin15a}
{Nayakshin} S.,  2015a, \mn@doi [\mnras] {10.1093/mnras/stu2074}, \href
  {http://adsabs.harvard.edu/abs/2015MNRAS.446..459N} {446, 459}

\bibitem[\protect\citeauthoryear{{Nayakshin}}{{Nayakshin}}{2015b}]{Nayakshin15b}
{Nayakshin} S.,  2015b, \mn@doi [\mnras] {10.1093/mnrasl/slu191}, \href
  {http://adsabs.harvard.edu/abs/2015MNRAS.448L..25N} {448, L25}

\bibitem[\protect\citeauthoryear{{Nayakshin}}{{Nayakshin}}{2015c}]{Nayakshin15c}
{Nayakshin} S.,  2015c, \mn@doi [\mnras] {10.1093/mnras/stv1915}, \href
  {http://adsabs.harvard.edu/abs/2015MNRAS.454...64N} {454, 64}

\bibitem[\protect\citeauthoryear{{Nayakshin}}{{Nayakshin}}{2016}]{Nayakshin16a}
{Nayakshin} S.,  2016, \mn@doi [\mnras] {10.1093/mnras/stw1404}, \href
  {http://adsabs.harvard.edu/abs/2016MNRAS.461.3194N} {461, 3194}

\bibitem[\protect\citeauthoryear{{Nayakshin}}{{Nayakshin}}{2017a}]{Nayakshin_Review}
{Nayakshin} S.,  2017a, \mn@doi [\pasa] {10.1017/pasa.2016.55}, \href
  {http://adsabs.harvard.edu/abs/2017PASA...34....2N} {34, e002}

\bibitem[\protect\citeauthoryear{{Nayakshin}}{{Nayakshin}}{2017b}]{Nayakshin17a}
{Nayakshin} S.,  2017b, \mn@doi [\mnras] {10.1093/mnras/stx1351}, \href
  {http://ukads.nottingham.ac.uk/abs/2017MNRAS.470.2387N} {470, 2387}

\bibitem[\protect\citeauthoryear{{Nayakshin} \& {Cha}}{{Nayakshin} \&
  {Cha}}{2012}]{NayakshinCha12}
{Nayakshin} S.,  {Cha} S.-H.,  2012, \mn@doi [\mnras]
  {10.1111/j.1365-2966.2012.21003.x}, \href
  {http://adsabs.harvard.edu/abs/2012MNRAS.423.2104N} {423, 2104}

\bibitem[\protect\citeauthoryear{{Nayakshin} \& {Fletcher}}{{Nayakshin} \&
  {Fletcher}}{2015}]{NayakshinFletcher15}
{Nayakshin} S.,  {Fletcher} M.,  2015, \mn@doi [\mnras]
  {10.1093/mnras/stv1354}, \href
  {http://adsabs.harvard.edu/abs/2015MNRAS.452.1654N} {452, 1654}

\bibitem[\protect\citeauthoryear{{Nayakshin} \& {Lodato}}{{Nayakshin} \&
  {Lodato}}{2012}]{NayakshinLodato12}
{Nayakshin} S.,  {Lodato} G.,  2012, \mn@doi [\mnras]
  {10.1111/j.1365-2966.2012.21612.x}, \href
  {http://adsabs.harvard.edu/abs/2012MNRAS.426...70N} {426, 70}

\bibitem[\protect\citeauthoryear{{Nayakshin}, {Dipierro}  \&
  {Szul{\'a}gyi}}{{Nayakshin} et~al.}{2019}]{NayakshinEtal19}
{Nayakshin} S.,  {Dipierro} G.,   {Szul{\'a}gyi} J.,  2019, \mn@doi [\mnras]
  {10.1093/mnrasl/slz087}, \href
  {https://ui.adsabs.harvard.edu/abs/2019MNRAS.tmpL..90N} {}

\bibitem[\protect\citeauthoryear{{Ndugu}, {Bitsch}  \& {Jurua}}{{Ndugu}
  et~al.}{2019}]{NduguEtal19}
{Ndugu} N.,  {Bitsch} B.,   {Jurua} E.,  2019, arXiv e-prints, \href
  {https://ui.adsabs.harvard.edu/abs/2019arXiv190611491N} {p. arXiv:1906.11491}

\bibitem[\protect\citeauthoryear{{Ormel} \& {Klahr}}{{Ormel} \&
  {Klahr}}{2010}]{OrmelKlahr10}
{Ormel} C.~W.,  {Klahr} H.~H.,  2010, \mn@doi [\aap]
  {10.1051/0004-6361/201014903}, \href
  {http://adsabs.harvard.edu/abs/2010A%26A...520A..43O} {520, A43}

\bibitem[\protect\citeauthoryear{{Paardekooper}, {Baruteau}, {Crida}  \&
  {Kley}}{{Paardekooper} et~al.}{2010}]{PaardekooperEtal10a}
{Paardekooper} S.-J.,  {Baruteau} C.,  {Crida} A.,   {Kley} W.,  2010, \mn@doi
  [\mnras] {10.1111/j.1365-2966.2009.15782.x}, \href
  {http://adsabs.harvard.edu/abs/2010MNRAS.401.1950P} {401, 1950}

\bibitem[\protect\citeauthoryear{{Pollack}, {Hubickyj}, {Bodenheimer},
  {Lissauer}, {Podolak}  \& {Greenzweig}}{{Pollack}
  et~al.}{1996}]{PollackEtal96}
{Pollack} J.~B.,  {Hubickyj} O.,  {Bodenheimer} P.,  {Lissauer} J.~J.,
  {Podolak} M.,   {Greenzweig} Y.,  1996, \mn@doi [Icarus]
  {10.1006/icar.1996.0190}, \href
  {http://adsabs.harvard.edu/abs/1996Icar..124...62P} {124, 62}

\bibitem[\protect\citeauthoryear{{Rafikov}}{{Rafikov}}{2005}]{Rafikov05}
{Rafikov} R.~R.,  2005, \mn@doi [\apjl] {10.1086/428899}, \href
  {http://ukads.nottingham.ac.uk/cgi-bin/nph-bib_query?bibcode=2005ApJ...621L..69R&db_key=AST}
  {621, L69}

\bibitem[\protect\citeauthoryear{{Rice}, {Lodato}  \& {Armitage}}{{Rice}
  et~al.}{2005}]{Rice05}
{Rice} W.~K.~M.,  {Lodato} G.,   {Armitage} P.~J.,  2005, \mn@doi [\mnras]
  {10.1111/j.1745-3933.2005.00105.x}, \href
  {http://ukads.nottingham.ac.uk/cgi-bin/nph-bib_query?bibcode=2005MNRAS.364L..56R&db_key=AST}
  {364, L56}

\bibitem[\protect\citeauthoryear{{Rosotti}, {Clarke}, {Manara}  \&
  {Facchini}}{{Rosotti} et~al.}{2017}]{RosottiEtal17}
{Rosotti} G.~P.,  {Clarke} C.~J.,  {Manara} C.~F.,   {Facchini} S.,  2017,
  \mn@doi [\mnras] {10.1093/mnras/stx595}, \href
  {https://ui.adsabs.harvard.edu/abs/2017MNRAS.468.1631R} {468, 1631}

\bibitem[\protect\citeauthoryear{{Safronov}}{{Safronov}}{1972}]{Safronov72}
{Safronov} V.~S.,  1972, {Evolution of the protoplanetary cloud and formation
  of the earth and planets.}.
Jerusalem (Israel): Israel Program for Scientific Translations, Keter
  Publishing House, 212 p.

\bibitem[\protect\citeauthoryear{{Santos} et~al.,}{{Santos}
  et~al.}{2017}]{SantosEtal17}
{Santos} N.~C.,  et~al., 2017, preprint, \href
  {http://adsabs.harvard.edu/abs/2017arXiv170506090S} {} (\mn@eprint {arXiv}
  {1705.06090})

\bibitem[\protect\citeauthoryear{{Schlaufman}}{{Schlaufman}}{2018}]{Schlaufman18}
{Schlaufman} K.~C.,  2018, \mn@doi [\apj] {10.3847/1538-4357/aa961c}, \href
  {http://adsabs.harvard.edu/abs/2018ApJ...853...37S} {853, 37}

\bibitem[\protect\citeauthoryear{{Shakura} \& {Sunyaev}}{{Shakura} \&
  {Sunyaev}}{1973}]{Shakura73}
{Shakura} N.~I.,  {Sunyaev} R.~A.,  1973, \aap, \href
  {http://cdsads.u-strasbg.fr/cgi-bin/nph-bib_query?bibcode=1973A%26A....24..337S&db_key=AST}
  {24, 337}

\bibitem[\protect\citeauthoryear{{Simon} et~al.,}{{Simon}
  et~al.}{2019}]{Simon19-ages}
{Simon} M.,  et~al., 2019, arXiv e-prints, \href
  {https://ui.adsabs.harvard.edu/abs/2019arXiv190810952S} {p. arXiv:1908.10952}

\bibitem[\protect\citeauthoryear{{Soderblom}, {Hillenbrand}, {Jeffries},
  {Mamajek}  \& {Naylor}}{{Soderblom} et~al.}{2014}]{SoderblomEtal14}
{Soderblom} D.~R.,  {Hillenbrand} L.~A.,  {Jeffries} R.~D.,  {Mamajek} E.~E.,
  {Naylor} T.,  2014, \mn@doi [Protostars and Planets VI]
  {10.2458/azu_uapress_9780816531240-ch010}, \href
  {https://ui.adsabs.harvard.edu/abs/2014prpl.conf..219S} {pp 219--241}

\bibitem[\protect\citeauthoryear{{Stamatellos}}{{Stamatellos}}{2015}]{Stamatellos15}
{Stamatellos} D.,  2015, \mn@doi [\apjl] {10.1088/2041-8205/810/1/L11}, \href
  {http://adsabs.harvard.edu/abs/2015ApJ...810L..11S} {810, L11}

\bibitem[\protect\citeauthoryear{{Thorngren}, {Fortney}, {Murray-Clay}  \&
  {Lopez}}{{Thorngren} et~al.}{2016}]{ThorngrenEtal15}
{Thorngren} D.~P.,  {Fortney} J.~J.,  {Murray-Clay} R.~A.,   {Lopez} E.~D.,
  2016, \mn@doi [\apj] {10.3847/0004-637X/831/1/64}, \href
  {http://adsabs.harvard.edu/abs/2016ApJ...831...64T} {831, 64}

\bibitem[\protect\citeauthoryear{{Vazan} \& {Helled}}{{Vazan} \&
  {Helled}}{2012}]{VazanHelled12}
{Vazan} A.,  {Helled} R.,  2012, \mn@doi [\apj] {10.1088/0004-637X/756/1/90},
  \href {http://adsabs.harvard.edu/abs/2012ApJ...756...90V} {756, 90}

\bibitem[\protect\citeauthoryear{{Vigan} et~al.,}{{Vigan}
  et~al.}{2017}]{ViganEtal17}
{Vigan} A.,  et~al., 2017, preprint, \href
  {http://adsabs.harvard.edu/abs/2017arXiv170305322V} {} (\mn@eprint {arXiv}
  {1703.05322})

\bibitem[\protect\citeauthoryear{{Vorobyov}}{{Vorobyov}}{2011}]{Vorobyov11}
{Vorobyov} E.~I.,  2011, \mn@doi [\apj] {10.1088/0004-637X/729/2/146}, \href
  {http://adsabs.harvard.edu/abs/2011ApJ...729..146V} {729, 146}

\bibitem[\protect\citeauthoryear{{Vorobyov} \& {Basu}}{{Vorobyov} \&
  {Basu}}{2005}]{VB05}
{Vorobyov} E.~I.,  {Basu} S.,  2005, \mn@doi [\apjl] {10.1086/498303}, \href
  {http://adsabs.harvard.edu/abs/2005ApJ...633L.137V} {633, L137}

\bibitem[\protect\citeauthoryear{{Vorobyov} \& {Basu}}{{Vorobyov} \&
  {Basu}}{2006}]{VB06}
{Vorobyov} E.~I.,  {Basu} S.,  2006, \mn@doi [\apj] {10.1086/507320}, \href
  {http://adsabs.harvard.edu/abs/2006ApJ...650..956V} {650, 956}

\bibitem[\protect\citeauthoryear{{Vorobyov} \& {Basu}}{{Vorobyov} \&
  {Basu}}{2010}]{VB10}
{Vorobyov} E.~I.,  {Basu} S.,  2010, \mn@doi [\apj]
  {10.1088/0004-637X/719/2/1896}, \href
  {http://adsabs.harvard.edu/abs/2010ApJ...719.1896V} {719, 1896}

\bibitem[\protect\citeauthoryear{{Weidenschilling}}{{Weidenschilling}}{1977}]{Weiden77}
{Weidenschilling} S.~J.,  1977, \mnras, \href
  {http://ukads.nottingham.ac.uk/abs/1977MNRAS.180...57W} {180, 57}

\bibitem[\protect\citeauthoryear{{Whipple}}{{Whipple}}{1972}]{Whipple72}
{Whipple} F.~L.,  1972, in {Elvius} A.,  ed., From Plasma to Planet. p.~211

\bibitem[\protect\citeauthoryear{{Williams} \& {Cieza}}{{Williams} \&
  {Cieza}}{2011}]{WilliamsCieza11}
{Williams} J.~P.,  {Cieza} L.~A.,  2011, preprint, \href
  {http://adsabs.harvard.edu/abs/2011arXiv1103.0556W} {} (\mn@eprint {arXiv}
  {1103.0556})

\bibitem[\protect\citeauthoryear{{Williams}, {Cieza}, {Hales}, {Ansdell},
  {Ruiz-Rodriguez}, {Casassus}, {Perez}  \& {Zurlo}}{{Williams}
  et~al.}{2019}]{WilliamsEtal19-ODISEA}
{Williams} J.~P.,  {Cieza} L.,  {Hales} A.,  {Ansdell} M.,  {Ruiz-Rodriguez}
  D.,  {Casassus} S.,  {Perez} S.,   {Zurlo} A.,  2019, \mn@doi [\apjl]
  {10.3847/2041-8213/ab1338}, \href
  {http://adsabs.harvard.edu/abs/2019ApJ...875L...9W} {875, L9}

\bibitem[\protect\citeauthoryear{{Woitke} et~al.,}{{Woitke}
  et~al.}{2019}]{WoitkeEtal19}
{Woitke} P.,  et~al., 2019, \mn@doi [\pasp] {10.1088/1538-3873/aaf4e5}, \href
  {https://ui.adsabs.harvard.edu/abs/2019PASP..131f4301W} {131, 064301}

\bibitem[\protect\citeauthoryear{{Zhang}, {Blake}  \& {Bergin}}{{Zhang}
  et~al.}{2015}]{ZhangEtal15-cond-front}
{Zhang} K.,  {Blake} G.~A.,   {Bergin} E.~A.,  2015, \mn@doi [\apjl]
  {10.1088/2041-8205/806/1/L7}, \href
  {http://adsabs.harvard.edu/abs/2015ApJ...806L...7Z} {806, L7}

\bibitem[\protect\citeauthoryear{{Zhang} et~al.,}{{Zhang}
  et~al.}{2018}]{Dsharp7}
{Zhang} S.,  et~al., 2018, \mn@doi [\apjl] {10.3847/2041-8213/aaf744}, \href
  {http://adsabs.harvard.edu/abs/2018ApJ...869L..47Z} {869, L47}

\bibitem[\protect\citeauthoryear{{Zhu}, {Hartmann}, {Nelson}  \&
  {Gammie}}{{Zhu} et~al.}{2012}]{ZhuEtal12a}
{Zhu} Z.,  {Hartmann} L.,  {Nelson} R.~P.,   {Gammie} C.~F.,  2012, \mn@doi
  [\apj] {10.1088/0004-637X/746/1/110}, \href
  {http://adsabs.harvard.edu/abs/2012ApJ...746..110Z} {746, 110}

\makeatother
\end{thebibliography}



\appendix

\section{The Dusty Disc model}\label{sec:DD}

\begin{figure*}
\includegraphics[width=0.99\textwidth]{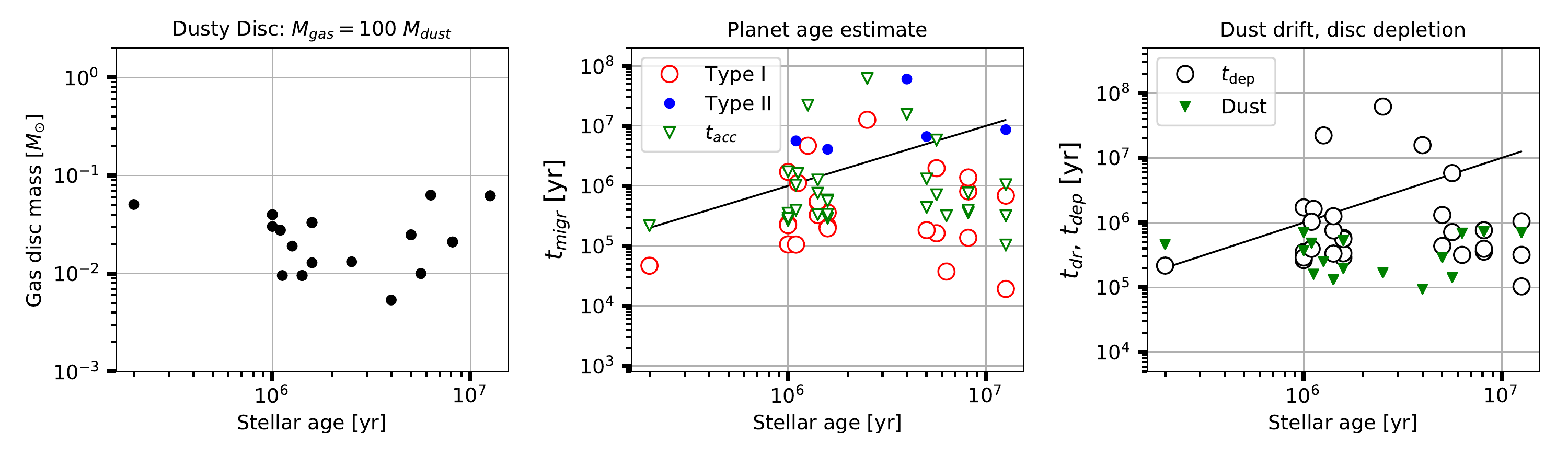}
\caption{Same as Fig. \ref{fig:SS} but for the steady state Dusty Disc scenario. The disc masses (left), the planet migration and accretion time scales (middle), the dust migration time and the disc depletion time (right). The evolutionary time scales are often shorter than the age of the systems, confirming the paradox of youth.}
\label{fig:DD}
\end{figure*}

In this section we show that ALMA planet candidates present the same `too young' challenge to the classical disc evolution paradigm in the {\bf Dusty Disc scenario} often used in the literature. In this case the gas disc mass is calculated by multiplying the dust disc mass by 100 for all the objects, $M_{\rm disc} = 100 M_{\rm dust}$, instead of using eq. \ref{Mdisc-ss0}. This is the only change to the approach taken in \S \ref{sec:CA}; the calculation of all of the time scales is then exactly the same. Further, now that the disc mass is not tied to the accretion rate onto the star, it is sensible to  define the disc depletion time as the time that it takes for it to be consumed by the star, $t_{\rm dep} = M_{\rm disc}/\dot M_*$.

Fig. \ref{fig:DD} shows the results for the sources for which we were able to find the dust disc masses in the literature. Comparing these results to Fig. \ref{fig:SS}, we see that tn the Dusty Disc scenario the planet migration and accretion time scales are somewhat longer, so the paradox of youth is not as acute, but it does remain a significant problem. Additionally, the right panel of fig. \ref{fig:DD} shows that with the smaller disc masses in the DD scenario, two additional problems arise. First, for many of the systems, these gas discs should have been consumed by the star by now. Second, the dust drift time are $\sim 10$ times shorter than $t_*$ for some of the systems. The DD model therefore suffers from three paradoxes of youth:  for the planets, the gas discs, and the mm-sized dust.

\section{Viscous equilibrium disc}\label{sec:SS-viscous}

\begin{figure*}
\includegraphics[width=0.99\textwidth]{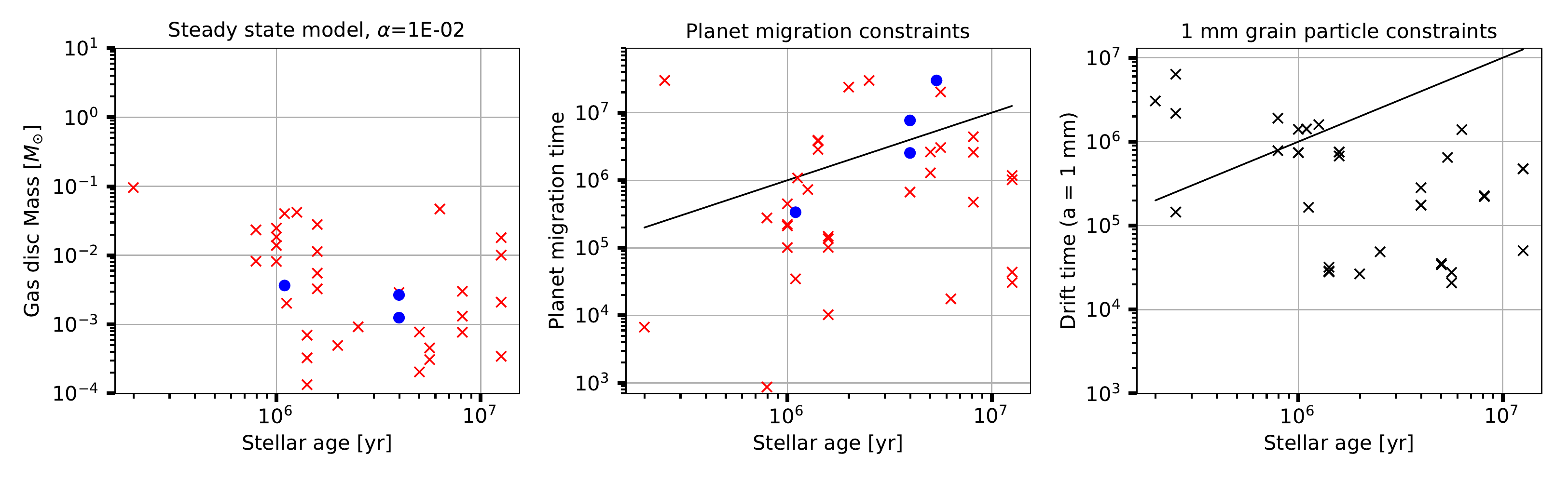}
\includegraphics[width=0.99\textwidth]{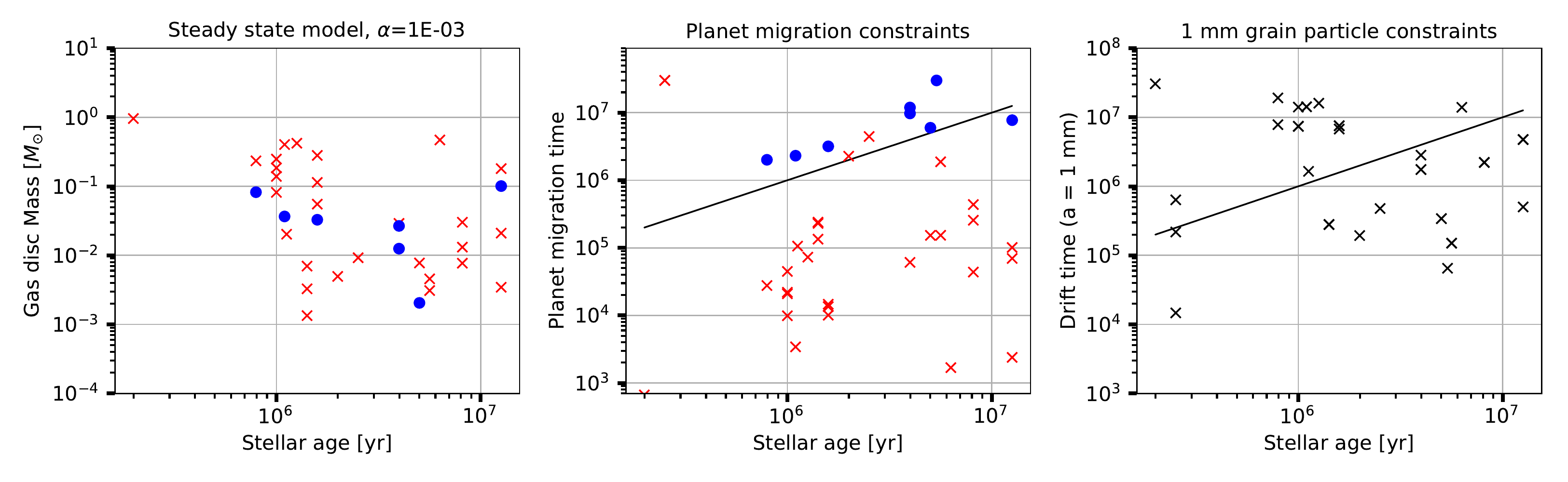}
\includegraphics[width=0.99\textwidth]{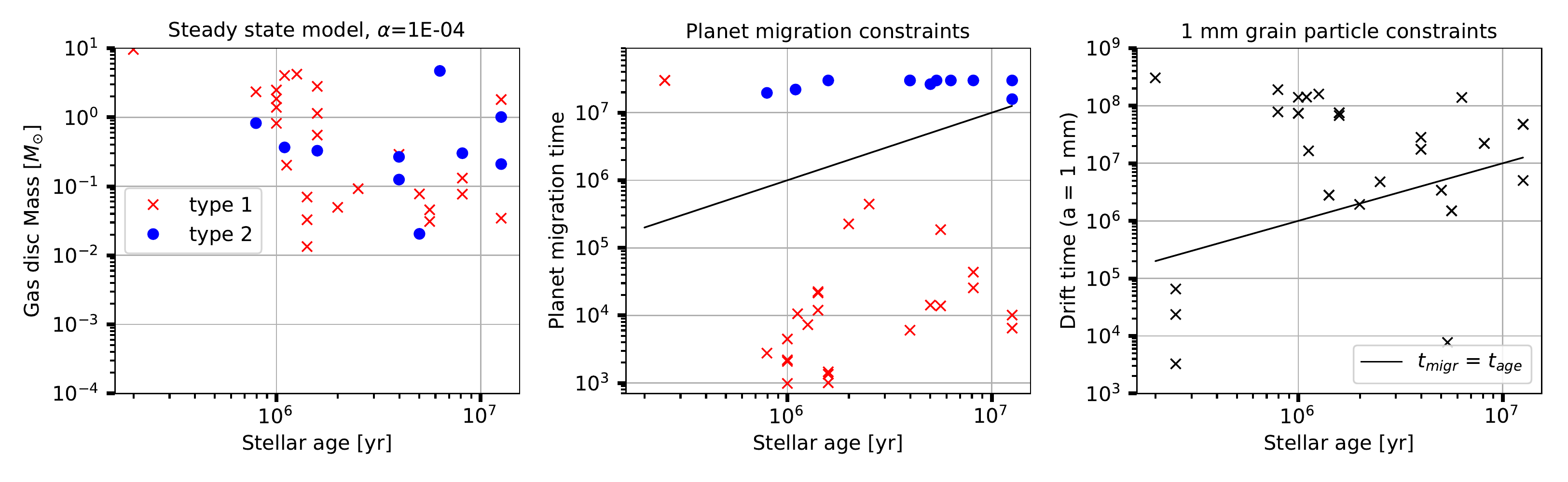}
\caption{The Steady State scenario for three different values of the viscoisty parameter, $\alpha_{\rm v} = 10^{-2}$, $10^{-3}$, $10^{-4}$, from the top to the bottom rows, respectively. See text in \S \ref{sec:SS-viscous} for detail.}
\label{fig:SS-different-alpha}
\end{figure*}

In \S \ref{sec:disc} we assumed that the disc is in viscous equilibrium and that the outer disc edge is at $R_{\rm out} = 200$~AU for all of our systems. We also fixed the viscosity parameter $\alpha_{\rm v} = 3\times 10^{-3}$. Here we relax these assumptions and instead let $\alpha_{\rm v}$ be a free parameter of the model.

Far from the stellar surface ($R \gg R_*$), the accretion rate through the part of the disc that has reached viscous equilibrium is constant with radius, and is related to the local disc properties via \citep{Shakura73}
\begin{equation}
    \dot M_* = 3\pi \nu \Sigma\;,
    \label{Mdot-0}
\end{equation}
where $\nu = \alpha_{\rm v} c_s H$ and $\Sigma$ are the disc viscosity and surface density, $c_s$ is the isothermal sound speed. Here we invert this equation to obtain $\Sigma$ at the location of the planet. This is the only change made to our approach.

Fig. \ref{fig:SS-different-alpha} shows the estimates for the total disc mass (left panel), the planet migration time scale (middle panel) and the dust drift time scales (right panel). The top, middle and the bottom rows differ by the value of the disc viscosity parameter, set to $\alpha_{\rm v} = 10^{-2}$, $10^{-3}$, $10^{-4}$, respectively. For the clarity of presentation the migration time is set $3\times 10^7$ yrs if it is longer than that (such migration times are long enough to not violate any constraints). The case with $\alpha_{\rm v} = 0.1$ is not presented here since for such a large value of $\alpha_{\rm v}$ the viscous time is, as shown previously, of order $10^{4} - 10^{5}$ years, far too short for the Steady State scenario since our systems are one-two orders of magnitude older than this.

As previously, the blue dots show candidate planets that are in the type 2 migration regime (a wide gap opened). At the largest value of $\alpha_{\rm v}$ considered in the figure (top row), very few of the planets are in this regime. The disc masses are moderate, with only a few systems found at mass $(0.01 -0.1)\msun$, with the rest being much smaller. The planet migration times are however still too short. In fact even the planets in the type 2 regime are "younger" than their stars because the viscous times are somewhat shorter than $t_*$ for $\alpha_{\rm v} =10^{-2}$. 

For $\alpha_{\rm v}=10^{-3}$, more planets are in type 2 regime, and these planets now are migrating sufficiently slow. The disc masses are however a factor of 10 higher than in the top row, so that many of the systems are now uncomfortable massive, e.g., $M_{\rm disc} \gtrsim 0.1\msun$. We expect such discs to show some signatures of spiral structure rather than cleanly defined rings. The planets migrating in the type 1 regime do so all too rapidly, typically in $\sim 1$\% of the system's age. 

These problems become even more apparent for the case of $\alpha_{\rm v} =10^{-4}$ (bottom row). Now most of the discs are massive enough to show spiral structure, and the type 1 migration times are {\bf very uncomfortably} short:  some planets have migration times $\lesssim 10^3$~yrs. The disc mass is however estimated here assuming the discs extend to 200 AU. The disc mass estimate could be reduced by a factor of several, typically, which would still be too high for the low $\alpha$ cases.

\section{Age uncertainties}\label{sec:age-un}

Due to a number of observational challenges and physical uncertainties in the evolution of young stars, their age determinations are thought to be reliable only for stars older than about 20 Myrs \citep{SoderblomEtal14}. For example, the exact nature of the energy dissipation at the accretion shock on the stellar surface can significantly influence the evolution of the stellar luminosity and radius and hence its inferred age \citep{BaraffeEtal12}. This usually make the stars appear older than they are. On the other hand, the presence of magnetic fields in young stars, usually neglected in the stellar evolution modelling of their ages, may lead to under-estimates of stellar ages by a factor of $\sim 2-3$ \citep{Simon19-ages}. 

\begin{figure*}
\includegraphics[width=0.33\textwidth]{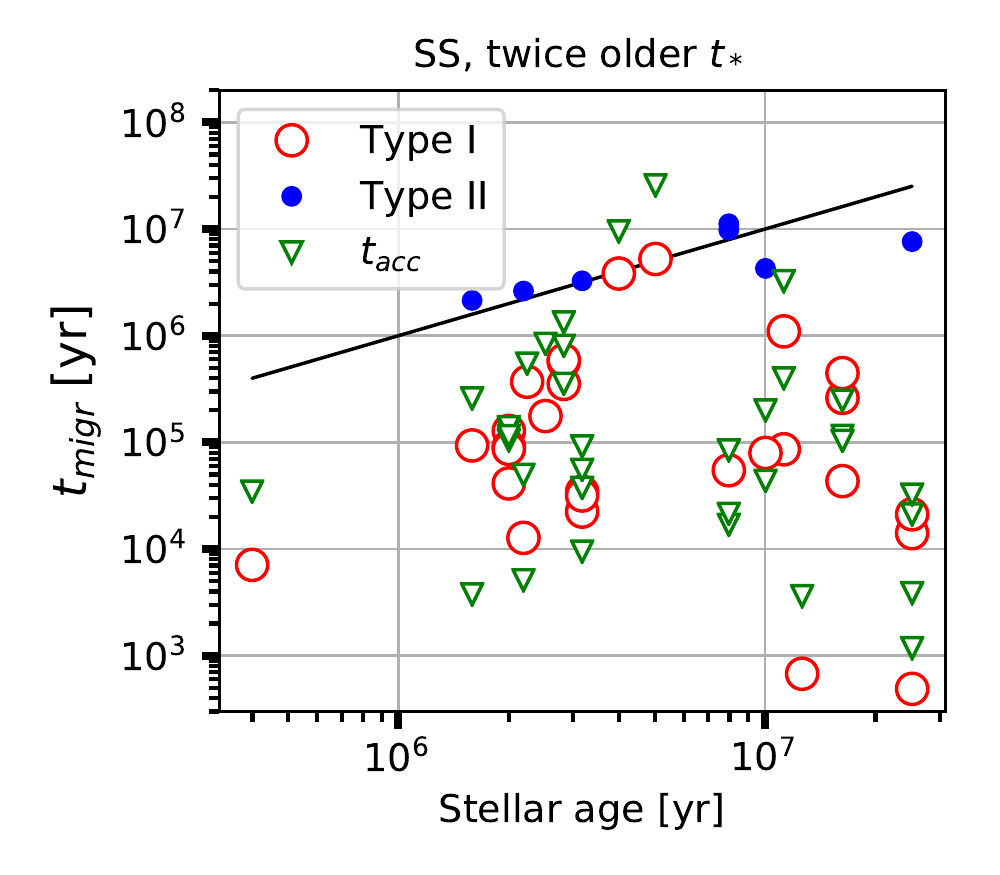}
\includegraphics[width=0.33\textwidth]{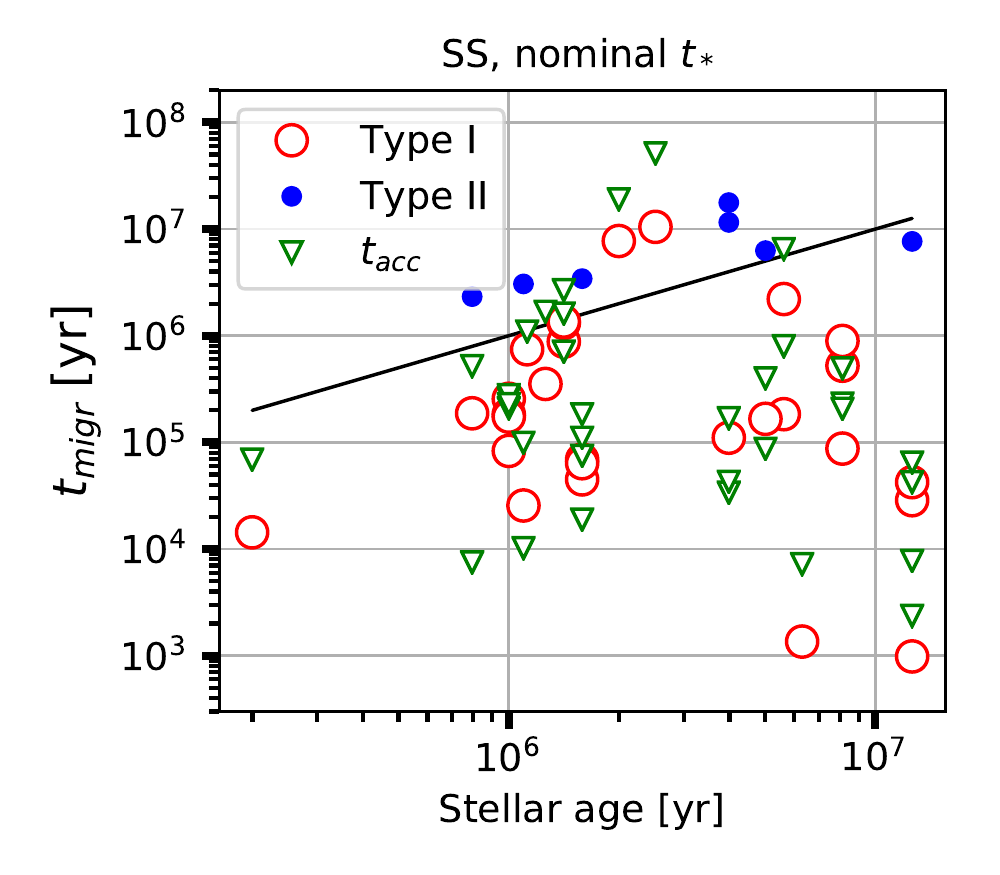}
\includegraphics[width=0.33\textwidth]{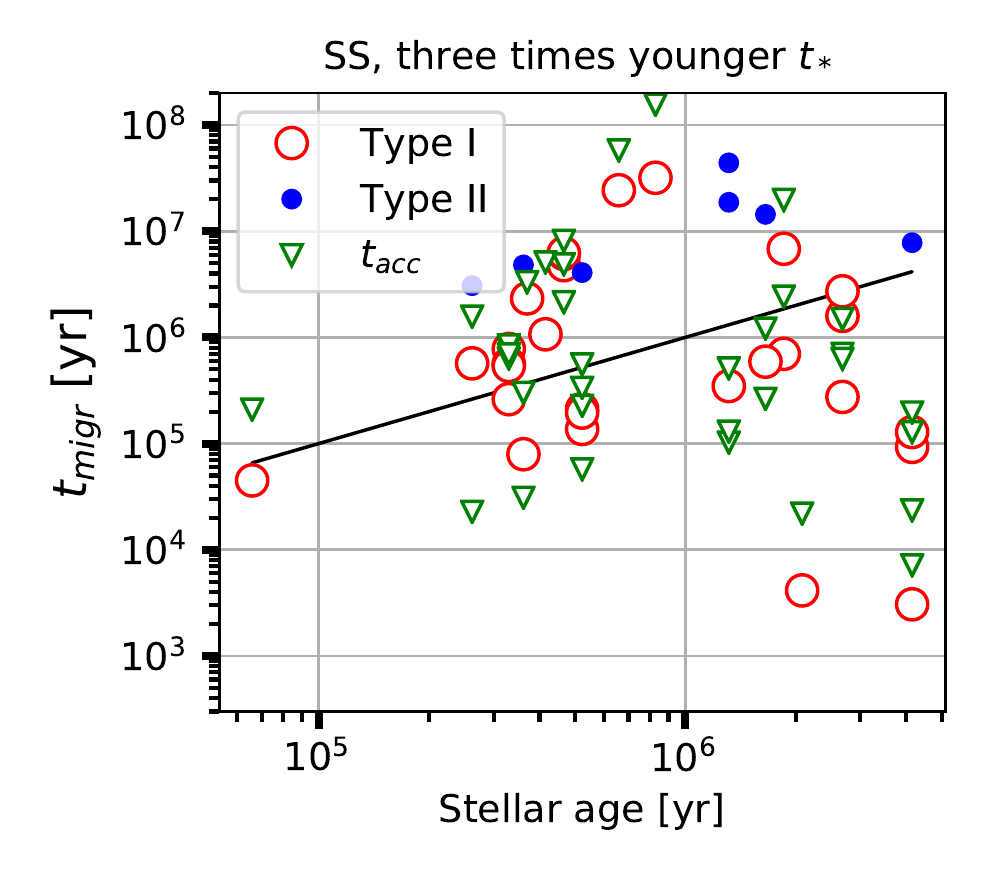}
\caption{{\bf Middle panel:} the migration and accretion time scales for the ALMA planet candidates as presented in fig. \ref{fig:SS}. {\bf Left panel:} Same calculation but assuming that all the stellar ages are twice longer. {\bf Right panel:} Same but assuming that all the stellar ages are three times younger.}
\label{fig:SS-different-ages}
\end{figure*}

To assess the importance of age uncertainties for our results, we repeat the analysis of \S \ref{sec:SS} assuming the stellar ages twice older (left panel of fig. \ref{fig:SS-different-ages}) or three times younger (right panel of the figure) than that currently inferred from observations. We can see that problems become more severe if stellar ages are under-estimated. This is caused by the disc mass estimate $M_{\rm disc} =\xi \dot M_* t_*$ increasing linearly with $t_*$. Both the planet migration and the gas runaway accretion timescales then become shorter as they are proportional to $1/t_*$. Therefore, the ratio of the accretion, and the type I migration, time scales to $t_*$ behave as $\propto t_*^{-2}$. This means that doubling $t_*$ makes the paradox of youth four times worse than for the nominal $t_*$ that we used in \S \ref{sec:SS}.

For the same reasons, decreasing $t_*$ by a factor of three relaxes the paradox by a factor of 9. We can see from the left panel of fig. \ref{fig:SS-different-ages} about a half of the objects now have migration and accretion time scales longer than the age estimate. However, there are still significant outliers for the older sources in particular. 

This shows that if the observed objects are significantly younger than usually assumed then the migration and accretion challenges to the standard scenario of planet formation and disc evolution disappear. However, the error in the age determination should be by as much as an order of magnitude for the older objects in particular, which appears unlikely. Further, taken in the broader context of protoplanetary disc evolution it would suggest that the discs are dispersed much faster than we currently assume. This would make the whole problem of planet formation by the Core Accretion scenario much more challenging. 

Therefore it appears unlikely that such significant stellar age over-estimates is the solution for the paradox of youth.



\bsp	
\label{lastpage}
\end{document}